\documentclass{aa}
\usepackage[varg]{txfonts}
\usepackage{graphicx}
\usepackage[colorlinks, urlcolor = blue]{hyperref}
\bibpunct{(}{)}{;}{a}{}{,} % to follow the A&A style

\usepackage[caption=false]{subfig}
\usepackage{gensymb}
\usepackage{textcomp}
\usepackage{ctable}
\usepackage[toc,page]{appendix}

\title{
	On the trail of a comet's tail: A particle tracking algorithm for comet \object{67P/Churyumov–Gerasimenko}
}

\author{
	Marius Pfeifer\inst{1}\thanks{Email: \href{mailto:pfeifer@mps.mpg.de}{pfeifer@mps.mpg.de}}
	\and Jessica Agarwal\inst{1, 2}
	\and Matthias Schröter\inst{3}
}

\institute{
	Max Planck Institute for Solar System Research, Justus-von-Liebig-Weg 3, D-37077 Göttingen, Germany
	\and Institute for Geophysics and Extraterrestrial Physics, TU Braunschweig, Mendelssohnstraße 3, D-38106 Braunschweig, Germany
	\and Max Planck Institute for Dynamics and Self-Organization, Am Faßberg 17, D-37077 Göttingen, Germany
}

\date{Received 2 August 2021 / Accepted 9 December 2021}

\abstract{
    During the post-perihelion phase of the European Space Agency's Rosetta mission to comet \object{67P}, the Optical, Spectroscopic, and Infrared Remote Imaging System on board the spacecraft took numerous image sequences of the near-nucleus coma, with many showing the motion of individual pieces of debris ejected from active surface areas into space.
}{
    We aim to track the motion of individual particles in these image sequences and derive their projected velocities and accelerations. This should help us to constrain their point of origin on the surface, understand the forces that influence their dynamics in the inner coma, and predict whether they will fall back to the surface or escape to interplanetary space.
}{
    We have developed an algorithm that tracks the motion of particles appearing as point sources in image sequences. Our algorithm employs a point source detection software to locate the particles and then exploits the image sequences' pair-nature to reconstruct the particle tracks and derive the projected velocities and accelerations. We also constrained the particle size from their brightness.
}{
    Our algorithm identified 2268 tracks in a sample image sequence. Manual inspection not only found that 1187 ($\sim$\,52\,\%) of them are likely genuine, but in combination with runs on simulated data it also revealed a simple criterion related to the completeness of a track to single out a large subset of the genuine tracks without the need for manual intervention. A tentative analysis of a small (\mbox{$n=89$}) group of particles exemplifies how our data can be used, and provides first results on the particles' velocity, acceleration, and radius distributions, which agree with previous work.
}

\keywords{
	comets: general 
	-- comets: individual: \object{67P/Churyumov–Gerasimenko} 
	-- zodiacal dust.
}

\begin{document}
	\maketitle

	\section{Introduction}

        Comets are relatively well-preserved remnant building blocks of our planets. Their interiors may provide us with clues about planetesimal formation and the composition of the outer solar nebula. One of the key quantities relevant in this context is the relative abundance of refractories and volatiles inside the cometary nucleus, often referred to as the refractory-to-ice (mass) ratio.
        
        This ratio cannot be measured directly with current spacecraft or remote observation techniques. An alternative is to determine it indirectly by measuring the dust-to-gas (mass) ratio of the material that was released from the nucleus into interplanetary space. This technique however comes with caveats. For one, estimating the lost dust mass relies on models that require knowledge of or assumptions about the dust size distribution and either optical properties (for remote sensing data) or spatial distribution (for in situ data). 
        
        Then, translating the dust-to-gas to the refractory-to-ice ratio is not straightforward. Part of the refractory material is contained in blocks that are too heavy to be accelerated past the comet's escape speed, and so they never leave the nucleus or fall back onto its surface \citep{choukroun-altwegg2020}. The blocks' volatiles on the other hand may have escaped entirely. Based on the coma dust-to-gas ratio, the refractory-to-ice ratio would thus be underestimated. Meanwhile, investigation of the surface would tend to overestimate it because of the refractory deposits. Such deposits may also quench the nucleus' outgassing \citep[e.g.,][]{gundlach-blum2016}, while ejected material may be outgassing too \citep{reach-vaubaillion2009}, making the translation to the refractory-to-ice ratio more complicated. Both issues would be better understood with a firmer grasp of the material's dynamics.
		
        The European Space Agency's Rosetta mission to comet \object{67P/Churyumov-Gerasimenko} has revealed that fall-back of refractory material is a common phenomenon, as wide parts of \object{67P}'s surface are uniformly covered in loose material \citep{thomas-davidsson2015}. Images obtained during the final descents of both the lander Philae \citep{mottola-arnold2015, pajola-mottola2016} and the Rosetta orbiter \citep{pajola-lucchetti2017} show the ground coated in a loose assembly of irregularly shaped blocks with typical sizes down to the centimeter-scale resolution limit of the images. Compared to the consolidated terrains thought to be more representative of the comet's "bedrock", from further out, these areas look relatively smooth. 
		
        Smooth terrains are predominantly found in the northern hemisphere \citep{thomas-davidsson2015}. This regional distribution of fall-back material is likely related to the asymmetric seasons on \object{67P}, with short ($\sim$\,1 year), hot perihelion summers in the southern hemisphere and long ($\sim$\,5.5 years), yet colder aphelion summers in the north \citep{keller-mottola2015_erosion}. \citet{pajola-lucchetti2017} have plausibly modeled the inter-region transport of fall-back material driven by the differences in local gas pressure due to varying solar irradiation. They showed that debris should be carried from regions of high gas pressure to regions where gas pressure is too low to keep the material afloat. \citet{lai-ip2016} have followed a more global approach studying the trajectories of dust particles embedded in a 3D Direct Simulation Monte Carlo model, and found that regional change in dust mantle thickness can be on the meter-scale.
        
        For such models, it is mandatory to have good knowledge of the debris' source distribution, (i.e., its production rate as a function of time and surface region), and the constituents' initial velocities and accelerations. But accelerations can also provide information about the ice content of larger chunks, because sublimating ice may manifest as an acceleration component toward the antisolar direction \citep{kelley-lindler2013, kelley-lindler2015, 10.1093/mnras/stw2179}.
        
        To learn more about these factors, \cite{10.1093/mnras/stw2179}  manually tracked 238 decimeter-sized particles in an image sequence obtained by the Optical, Spectroscopic, and Infrared Remote Imaging System (OSIRIS) on January 6, 2016. Of the particles whose projected velocities were measured, at least 10\,\% were faster than the local escape speed of the nucleus; hence they likely reached interplanetary space, contributing to the comet's debris trail \citep[e.g.,][]{sykes-walker1992a}. \citet{keller-mottola2017} estimated that of the remaining 90\,\%, at least 20\,\% fell back to the surface within several hours, possibly accumulating in a regional layer of debris that still contains some water ice.
    	
    	Other studies that looked for ejected debris in OSIRIS images also did so manually, or mostly by making use of the elongated particle trails in long-exposure images: \citet{bertiniSearchSatellitesComet2015} searched for satellites near \object{67P} using the SExtractor software \citep{1996A&AS..117..393B}, but found no unambiguous candidates. A moon (dubbed "churymoon") was later discovered visually by \citet{landru79MaybeLittleCoorbital2019}, and in the following tracked by \citet{marin-yaselidelaparraAnalysisParticleSurroundings2020} using TrackMate \citep{tinevezTrackMateOpenExtensible2017}. \citet{rotundiDustMeasurementsComa2015} and \citet{fulleEvolutionDustSize2016} manually identified $\sim$\,400 and 204 particles respectively using image differencing. \citet{davidssonOrbitalElementsMaterial2015} detected, manually tracked, and determined the orbital elements of four particles. \citet{ottPaDeParticleDetection2016} developed an algorithm to detect elongated particle trails based on Canny edge detection \citep{cannyComputationalApproachEdge1986} and Hough transformation \citep{houghMachineAnalysisBubble1959}, and used it to measure 262 particles \citep{drolshagenDistanceDeterminationMethod2017, ottDustMassDistribution2017}. \citet{guttlerCharacterizationDustAggregates2017} used the blurriness of defocussed particles--instead of the parallax effect used in most of the previous studies--to derive the properties of 109 particles semi-automatically. Finally, \citet{frattinPostperihelionPhotometryDust2017} developed an automated detection method based on line-shaped matching functions to detect elongated particle trails, and identified 1925 tracks (and again 1916 tracks in \citealp{frattinObservationalConstraintsDynamics2021}). No algorithm however has been developed for OSIRIS images to automatically track point-source-like particles.
        
        During the post-perihelion phase of the mission, OSIRIS has regularly obtained image sequences like the one analyzed by \citet{10.1093/mnras/stw2179}. Many of them show fountains of debris that seem to stem from locally confined sources. Exploiting the sequences' very specific properties, we have created a tool that can automatically detect and track the motion of the point-source-like debris.
    
    	Instead of tracing particle tracks manually on a stacked image sequence as performed by \citet{10.1093/mnras/stw2179}, our algorithm first examines each image individually before recovering tracks from the gathered data. Because the particles scatter sunlight but are not spatially resolved, they appear as point sources in the images, which already distinguishes them from most other features. Nevertheless we further clean the images to improve their signal-to-noise ratio and then use the SExtractor software \citep{1996A&AS..117..393B} to detect them. Once located, their positions are passed on to the core of our project, the tracking algorithm. Here, we exploit the dataset's pair-nature to reconstruct the particle motions. 
    	
    	Our work presents a new approach in the large field of particle tracking \citep[for other disciplines see e.g.,][]{westerweelParticleImageVelocimetry2013, chenouardObjectiveComparisonParticle2014, ulmanObjectiveComparisonCelltracking2017, roseParticleTrackingNanoparticles2020}. In astronomy, tracking algorithms were developed to discover small Solar System bodies (SSSBs) in large-scale sky surveys, such as: Spacewatch \citep[MODP][]{1991AJ....101.1518R}; CFHT, SKADS, OSSOS \citep[MOP][]{10.1111/j.1365-2966.2004.07217.x, GLADMAN2009104, Bannister2016}; Pan-STARRS, LSST \citep[MOPS][]{KUBICA2007151, Denneau2013, JONES2018181}; (NEO)WISE \citep[WMOPS][]{Mainzer_2011}; PTF \citep[PTF MOPS][]{10.1093/mnras/stt951}; and ZTF \citep[ZMODE][]{Masci2018}. Recently, also more general-purpose SSSB discovery engines have been developed, such as HelioLinC \citep{Holman2018}, THOR \citep{Moeyens2021}, or tracee \citep{ohsawaDevelopmentTrackletExtraction2021}. All these algorithms were however mostly designed for Earth-based observations of SSSBs, where the object density is low, the apparent speed small, and the motion near linear (as pointed out by \citealp{https://doi.org/10.1029/2019EA000843}). They are therefore not very well suited for the dense and more dynamic dust environment in the coma of \object{67P}.
    	
    	Much closer related to our project are the methods for particle tracking around asteroid \object{(101955) Bennu}, which were developed in parallel to our own algorithm and recently published by the OSIRIS-REx team (see \citealp{https://doi.org/10.1029/2020JE006549} for an overview of the special issue). Following the discovery of \object{Bennu}'s activity \citep{theosiris-rexteamOperationalEnvironmentRotational2019, doi:10.1126/science.aay3544}, \citet{https://doi.org/10.1029/2019EA000843} developed a dedicated algorithm to detect and track the ejected material. \citet{https://doi.org/10.1029/2019EA000938},  \citet{https://doi.org/10.1029/2019EA000937}, and \citet{https://doi.org/10.1029/2019JE006363} then estimated the trajectories and orbits of the identified particles and traced them back onto the surface of \object{Bennu} to reconstruct the ejection events.
		
        In this paper we describe our methodology in detail and apply it to the same image sequence analyzed by \citet{10.1093/mnras/stw2179}. Because our algorithm can detect much fainter tracks, we find more than three times as many tracks than the manual procedure and hence significantly improve the statistics. 
	    		
		\begin{figure}[!b]
  			\centering
			\includegraphics[width=\linewidth]{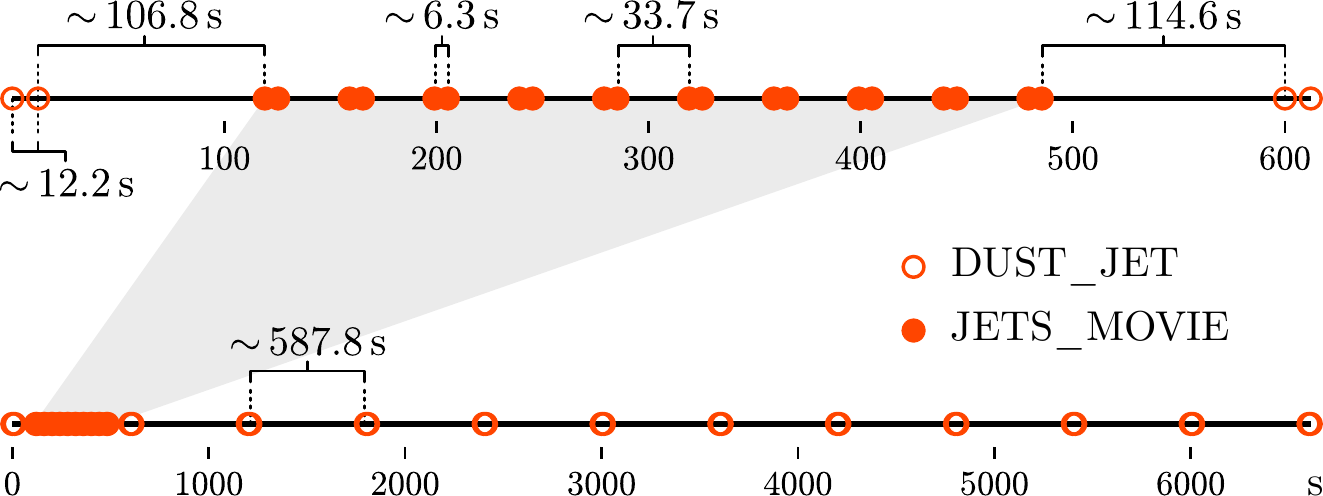}
			\caption{Timeline of image sequence STP090. The sequence was constructed from the two subsequences "JETS\_MOVIE" (20 images, principal sequence) and "DUST\_JET" (24 images, extended sequence). The whole sequence spans almost two hours. Due to the alternating time intervals between recordings, the images come in pairs.}
			\label{fig:timeline}
		\end{figure}
		
		\begin{figure*}[!ht]
  			\centering
			\includegraphics[width=\linewidth]{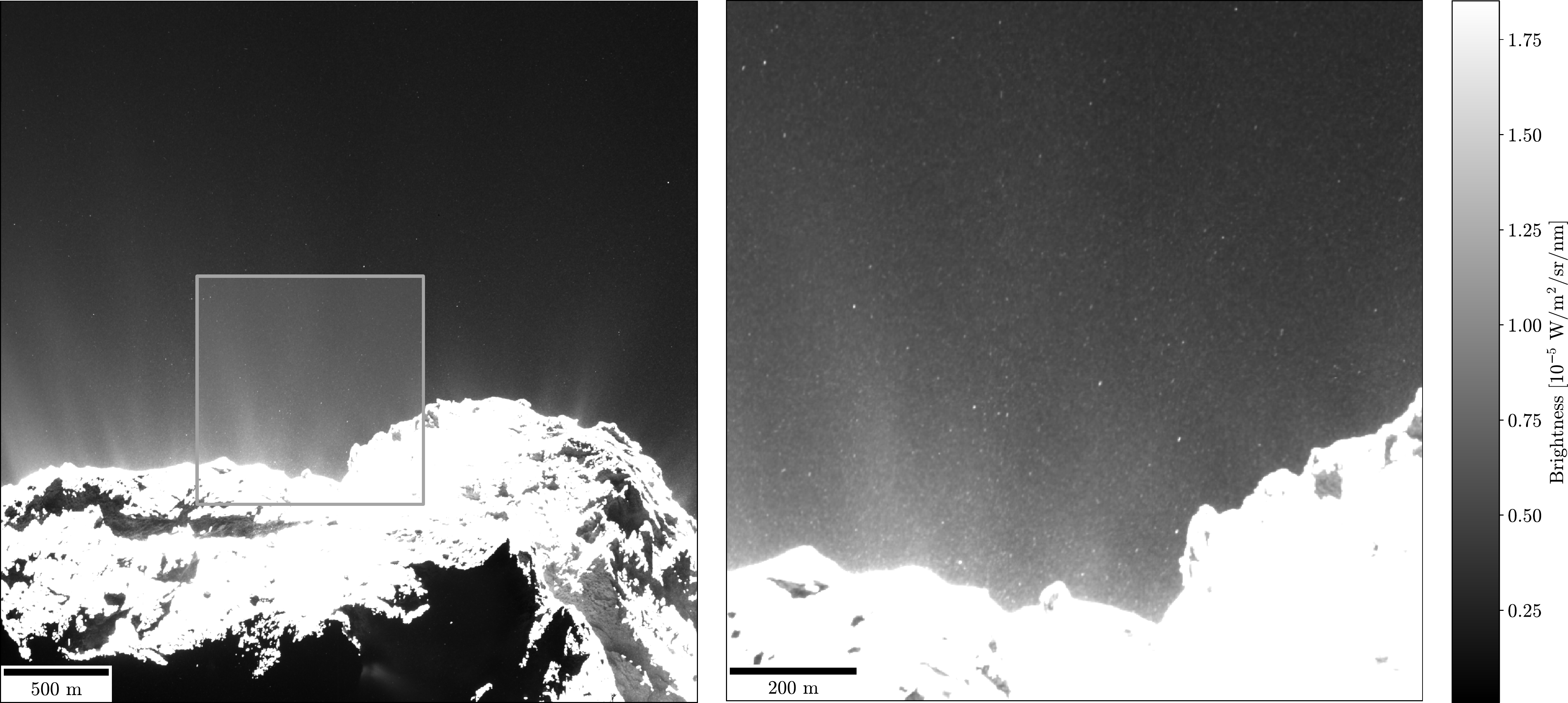}
            \caption{One of the source images of STP090. It showcases the appearance of dust particles as small point sources, as well as the bright surface of the irradiated nucleus and its radiant features. Full image on the left, close-up of central region on the right (contrast enhanced for visibility).}
            \label{fig:source}
        \end{figure*} 
        
        The data and image processing are described in Sect.~\ref{sec:data}. The tracking algorithm is described in Sect.~\ref{sec:tracking}, and the parameter optimization in Sect.~\ref{sec:optimization}. In Sect.~\ref{sec:results}, we evaluate the performance of the algorithm and present initial scientific results. Our findings are summarized in Sect.~\ref{sec:conclusions}. The entire project is written in the Python programming language version 3.7.7 \citep{van1995python} and we are happy to provide access to the code on request.

	\section{Data}
	\label{sec:data}
		
	    The source material for our tracking algorithm are the previously discussed image sequences recorded by OSIRIS' Narrow Angle Camera (NAC) on board the Rosetta spacecraft (for NAC specifications see Table~\ref{tab:mission_details} and \citealp{2007SSRv..128..433K}). The sequences typically show (parts of) \object{67P}'s nucleus and coma from $\sim$\,20--400\,km distance, and share a characteristic that is essential for our tracking algorithm: their images were recorded in pairs with the time interval between pairs being much longer than the intra-pair cadence (see Sect.~\ref{sect:pair_tracking}).

        We refer to the image sequence that we exemplarily analyze in the following as STP090. It was obtained with NAC on January 6, 2016, starting from UT 07:01:03 when Rosetta was $\sim$\,86\,km away from the nucleus, and \object{67P} was at a heliocentric distance of $\sim$\,2.06\,AU post-perihelion. We constructed STP090 from two sub-sequences, a short and a long one. The short sequence (OSIRIS activity tag "JETS\_MOVIE") consists of 20 images and covers roughly six minutes, while the long sequence (OSIRIS activity tag "DUST\_JET") contains 24 images and spans almost two hours, starting roughly two minutes before the short one (see Fig.~\ref{fig:timeline}). While the exposure time for the short sequence is constant at 0.24\,s, it alternates between 0.24 and 6\,s for the long one. In the following we refer to the short and long one as the principal and extended sequences respectively, a distinction that becomes clearer in Sect.~\ref{sect:extended_tracking}. The relevant mission details are summarized in Table~\ref{tab:mission_details}.
        
        \ctable[
            caption = Mission details for sequence STP090.,
            label   = tab:mission_details,
            pos     = b,
            width   = \linewidth,
            notespar
        ]{lc}{
        }{                                                                          \FL
        	Date of recording      & January 6, 2016   \NN 
        	Time of recording      & UT 07:01:03--UT 08:51:15              \NN
        	Total duration         & 1\,h 50\,min 11\,s                          \NN
        	Heliocentric distance  & $\sim 2.06$\,AU                       \NN
        	Nucleocentric distance    & $\sim 86$\,km                         \ML
        	Camera                 & OSIRIS NAC                            \NN
        	Field of view (FOV)    & $2.208\degree \times 2.208\degree$    \NN 
        	CCD resolution         & $2048 \times 2048$\,px \hspace{6pt}|\hspace{6pt} $\sim 3.3 \times 3.3$\,km  \NN
        	Pixel resolution       & $18.6 \times 18.6$\,\textmu rd \hspace{6pt}|\hspace{6pt} $\sim 1.6 \times 1.6$\,m   \NN
        	Filter (NAC F22)       & center: 649.2\,nm, bandwidth: 85\,nm \LL
        }

        We use OSIRIS images of calibration level 3E (Committee on Data Management, Archiving, and Computing, CODMAC, level 4), which includes solar and in-field stray-light correction, radiometric calibration and geometric distortion correction\footnote{The data are available at the Planetary Science Archive of the European Space Agency under \url{https://www.cosmos.esa.int/web/psa/rosetta}.} \citep{2015A&A...583A..46T}. The pixel values of this level are provided in radiance units (W/m$^2$/sr/nm).

        At this "raw" stage, it is already possible to see some of the brighter particles (see Fig.~\ref{fig:source}). To also detect fainter particles however and track their motion, the images are first cleaned before the point source coordinates are extracted (see Sect.~\ref{sec:detection}).

		\subsection{Image cleaning}
		\label{sec:cleaning}
		
    		\begin{figure}[!ht]
      			\centering
    			\includegraphics[width=\linewidth]{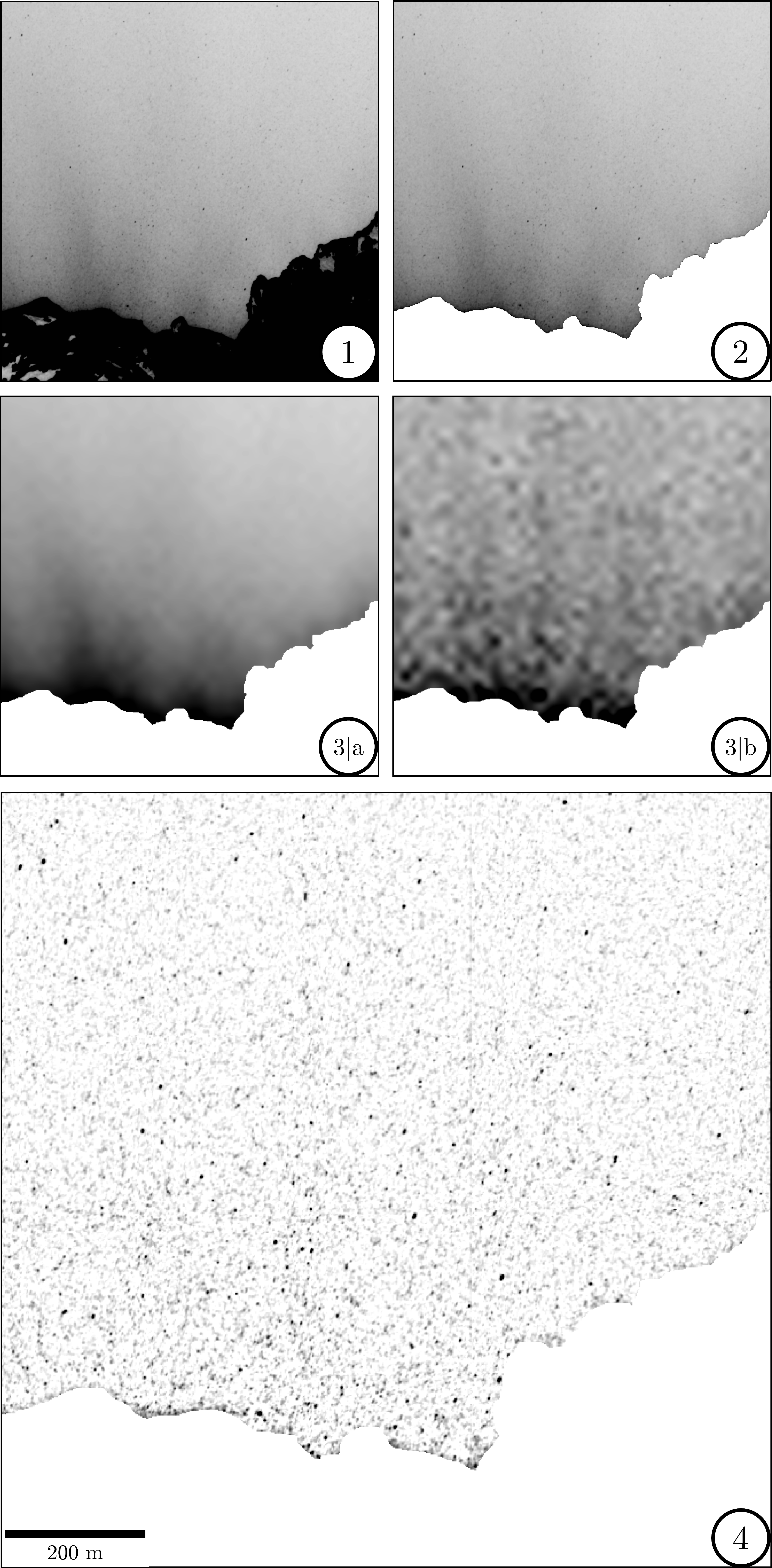}
    			\caption{Illustration of the cleaning pipeline: (1) the unaltered source image (OSIRIS level 3E); (2) the masked-out nucleus; (3) the estimated background level (a) and the corresponding RMS map of the background-subtracted image (b); (4) the background- and nucleus-subtracted image predominated by dust particles. All images show the same central region indicated in Fig.~\ref{fig:source} and are brightness-inverted for better reading.}
    			\label{fig:cleaning}
    		\end{figure}

			To optimize particle detection, we aim to minimize signals not associated with point sources. Of those, we identify three types: (1) ambient background noise that stems from the diffuse coma and bright, roughly cone-shaped dust streams radiating from the nucleus (in the following called radiant features, Fig.~\ref{fig:source}); (2) prints of cosmic ray hits; and (3) the nucleus itself. While cosmic ray hits may confuse the point source detector occasionally, we found that due to their small number, they do not significantly affect the tracking results. The background noise and the nucleus however need to be removed.
						
    		\begin{figure*}[!ht]
      			\centering
    			\includegraphics[width=\linewidth]{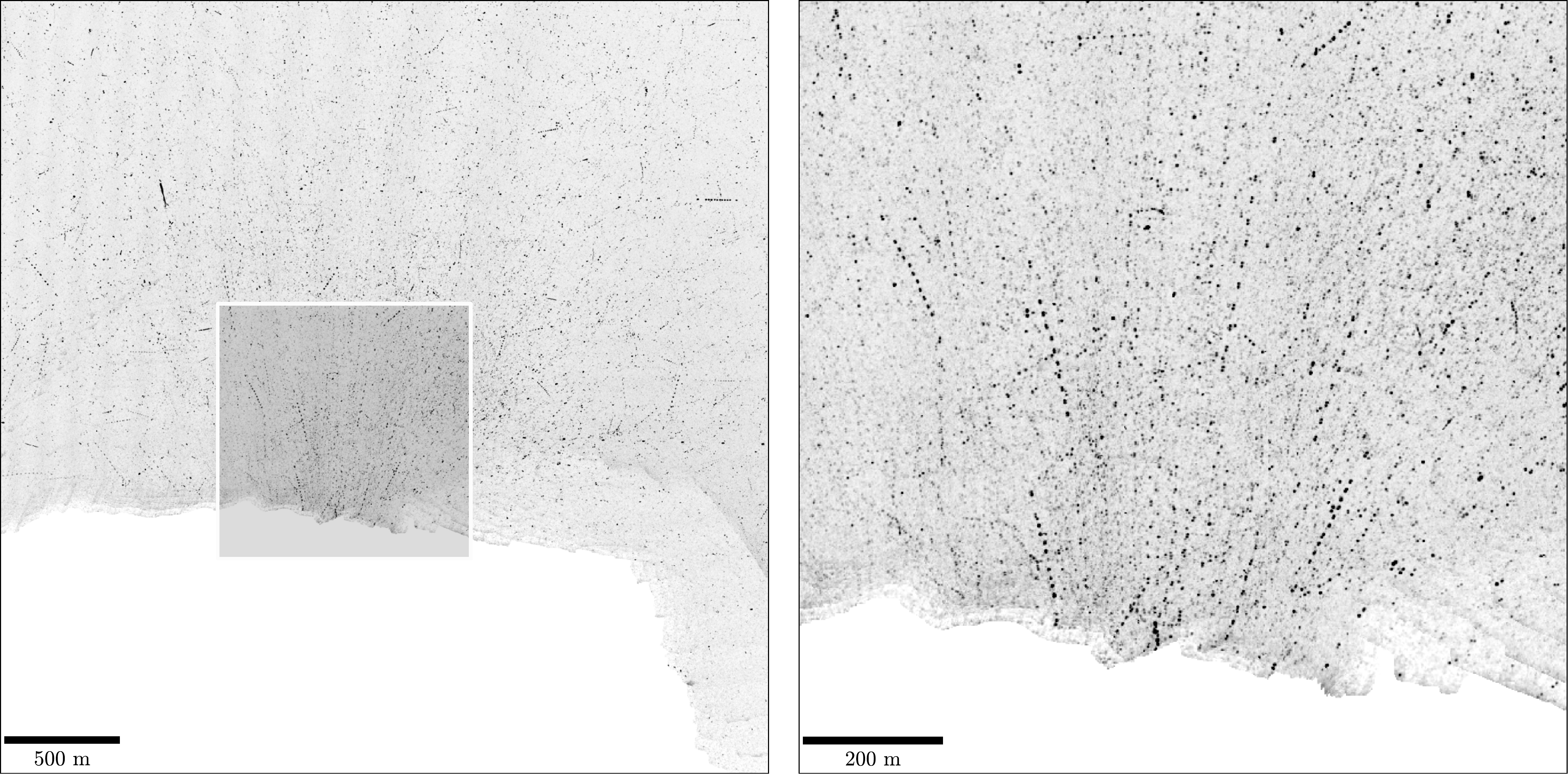}
    			\caption{Stacked image of sequence STP090. It was created by selecting the maximum value for each pixel across the image sequence. Full (brightness-inverted) image on the left, close-up of central region on the right. The stacked image is only used to check the tracking results and identify sidereal objects.}
    			\label{fig:stack}
    		\end{figure*}
		
    		\begin{figure*}[!ht]
      			\centering
    			\includegraphics[width=\linewidth]{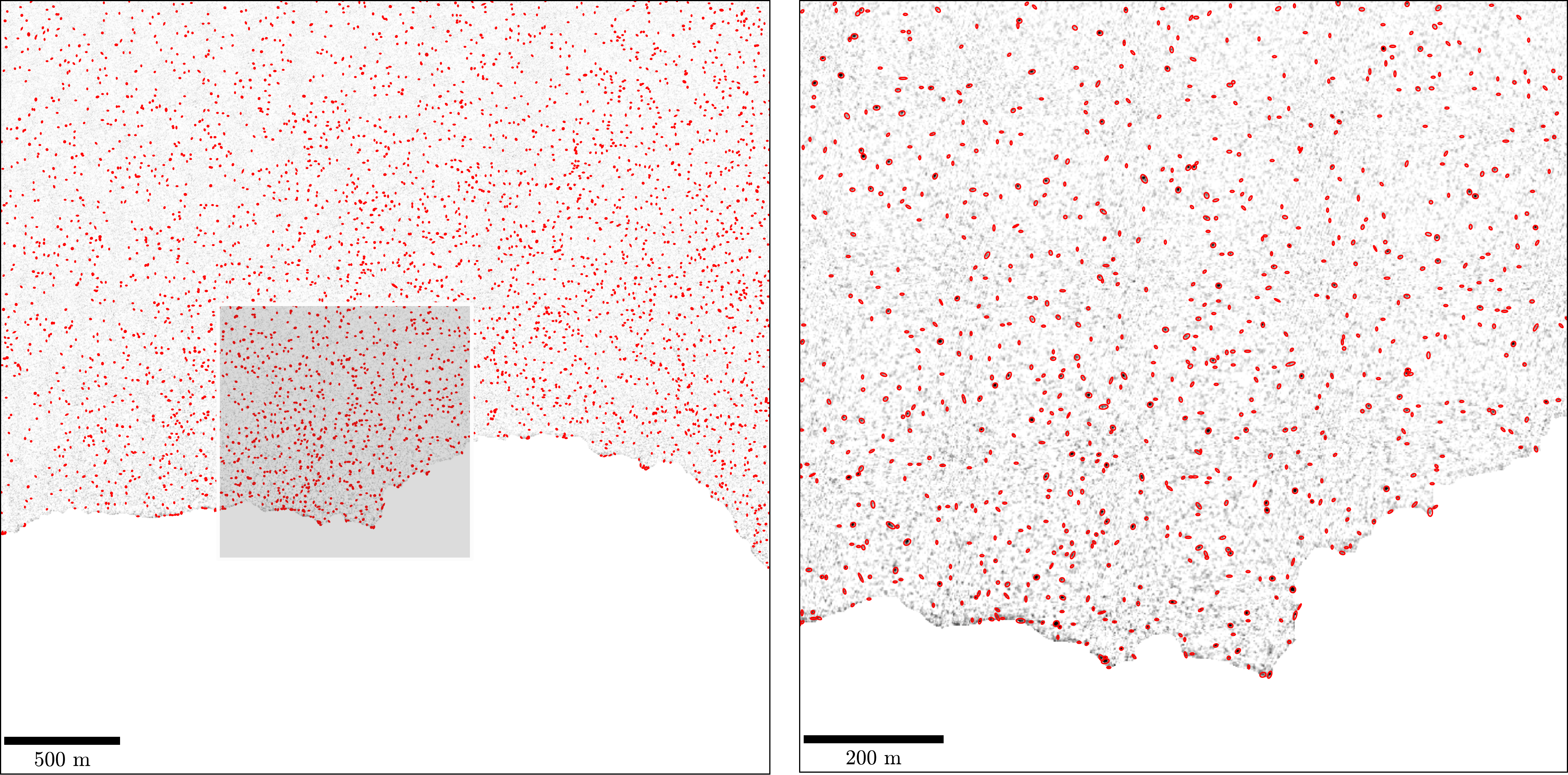}
    			\caption{Sample detection set (red ellipses) from one of the processed images of sequence STP090. Full image on the left, close-up of central region on the right.}
    			\label{fig:detections}
    		\end{figure*}
			
			The diffuse coma signal is determined by the background estimator from the library for Source Extraction and Photometry \citep[SEP,][]{Barbary2016, 1996A&AS..117..393B, 1987PASP...99..191S}. It subdivides an image into a grid of rectangular sections, calculates the background locally for each (with the help of iterative $\kappa$-$\sigma$-clipping and mode estimation), and merges the resulting background patches smoothly back together (via natural bicubic spline interpolation) to form the global background map. This approach has the advantage that it can account for medium-scale changes in the background level--such as radiant features--and is therefore generally well-suited for our datasets (Fig.~\ref{fig:cleaning}). 
			
			The bright nucleus on the other hand poses an issue for the background estimation. For sections at its limb that include both nucleus and coma the background level would be overestimated. To prevent this, we mask out the entire nucleus using its approximate shape retrieved from the OSIRIS level 4S (CODMAC level 5) georeferencing layers, and subsequently refine the mask with the help of edge-detection algorithms. The shape is then passed on to the background estimator which ignores the masked area during processing. Because particle detection is not possible in front of the illuminated surface, by removing the nucleus we only lose information of particles that appear in front shadowed regions.
			
			The background estimation also renders a root mean square (RMS) map. It is calculated in a similar fashion as the background signal, where the RMS values are first determined locally, before being smoothed out to form the global map. Since the RMS map is calculated from the background-subtracted image, it gives us an idea about the remaining random noise. This information is used during the point source detection.
			
			With the nucleus and background removed, the images are predominated by the signal of dust particles. We call the remaining area that still contains data the dust field. 
			
			Lastly, the processed images are stacked by selecting the maximum value that each pixel assumed over the sequence (see Fig.~\ref{fig:stack}). Unlike \cite{10.1093/mnras/stw2179} however, we do not use this stacked image to track the particles, but instead only as a visual aid to check the tracking results and identify sidereal objects.

		\subsection{Point source detection}
		\label{sec:detection}
	
			The particles are detected with the help of the SEP software \citep{Barbary2016, 1996A&AS..117..393B}. It employs a thresholding approach based on \citet{1980CompJ..23..262L}'s one-pass algorithm that can be used to identify point sources. Only pixels whose values are above the local RMS level (see Fig.~\ref{fig:cleaning}) multiplied by some user-defined detection threshold are considered during the detection process. 
			
			The algorithm then extracts sources based on the number of contiguous pixels, which are later-on deblended (using their brightness topology, \citealp{1990MNRAS.247..311B}) to separate neighboring point sources that have been extracted together. The resulting dataset can additionally be "cleaned", meaning it is checked whether each source would have also been detected without its neighbors being present. In the following we call identified sources detections, and the entirety of all sources detected in a single image a detection set. Figure~\ref{fig:detections} shows a sample of such a set.

	\section{Tracking algorithm}
	\label{sec:tracking}

		In the following we assume an image sequence comprised of $N$ images recorded at times $t_n$ ($n \in \{0, 1, ..., N-1\}$), thus containing $N/2$ image pairs. We start by briefly defining key concepts:
		
		\begin{description}
		    \item[Track and candidate track:] Track refers to a collection of detections that are all of the same object and thus depict the object's path through the recorded scene. Candidate track refers to any collection of detections, independent of whether or not they belong to the same object. They are only accepted as tracks once they pass a quality check.
		    \item[Pursuit and tracking run:] Pursuit refers to the tracking of a single object throughout a dataset, while tracking run describes the exhaustive analysis of an entire dataset, encompassing every possible pursuit for a fixed set of tracking parameters.
		    \item[Tracking parameters:] Tracking parameters are the parameters that govern the execution of a tracking run and each of its pursuits. They influence for example which detections and detection pairs are considered during a pursuit and define how many of them candidate tracks must contain to be accepted. 
		\end{description}

		\subsection{Pair-tracking}
		\label{sect:pair_tracking}
				
			\begin{figure}[!ht]
	  			\centering
				\includegraphics[width=\linewidth]{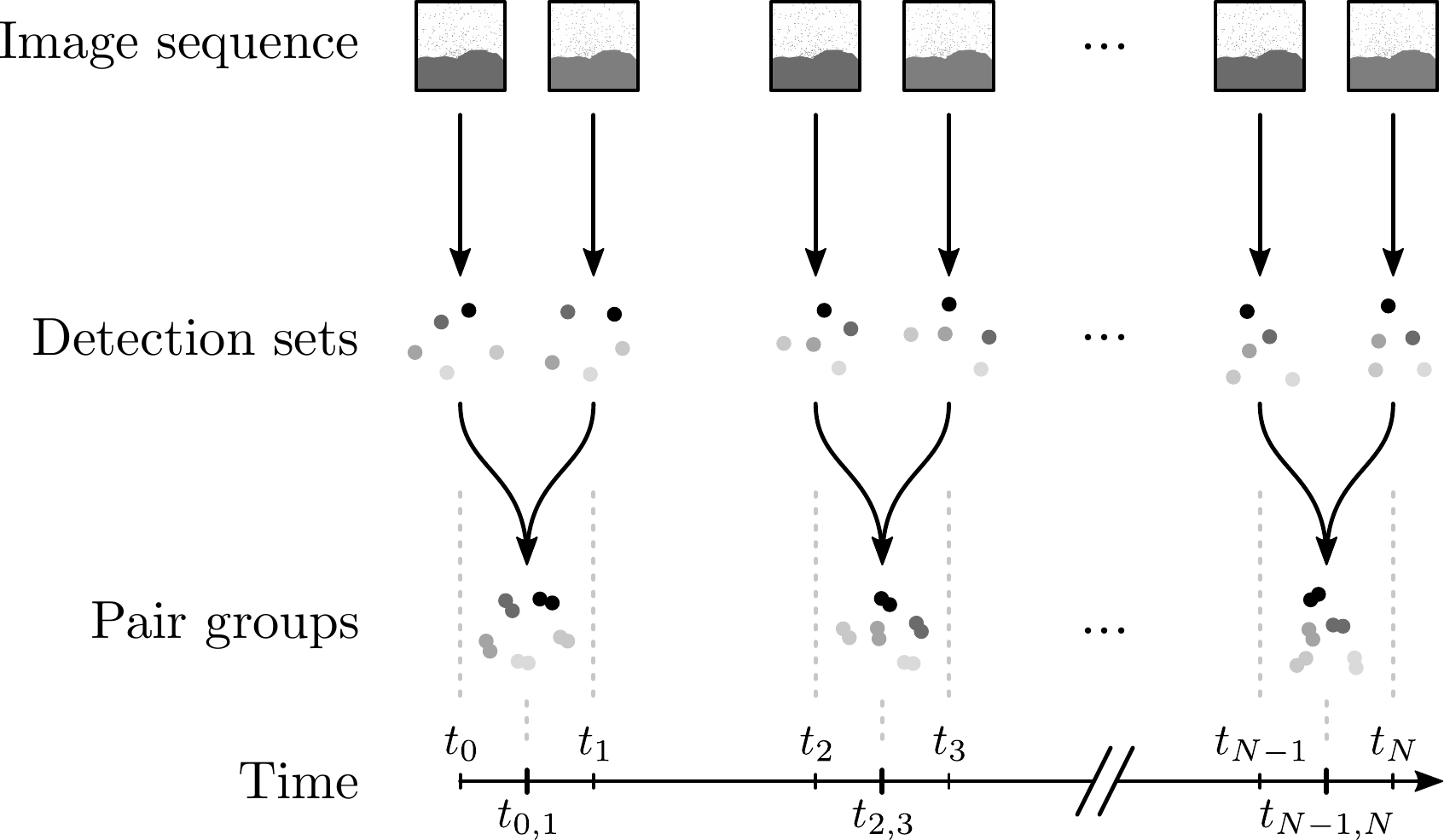}
				\caption{Illustration of the relation between image sequence, detection sets, and pair groups. Together, the pair groups comprise the pool of available pairs.}
				\label{fig:data_reduction}
			\end{figure}
			
			A central aspect of our tracking algorithm is the exploit of the image sequences' pair-nature. We assume that during the time interval of an image pair, the particles only travel a short distance (typically no more than a few pixels). This allows us to pair neighboring detections, one from each of the images, and analyze them as a unit: close detections likely belong to the same object. Consequently, our tracking algorithm predominantly operates pairwise, reverting to search for single detections only when there are no suitable pairs. We call this process pair-tracking.

			To create the detection pairs, our algorithm iterates over the detection sets of each image pair. For every detection in the first set, the algorithm looks for detections in the second set within a predefined search radius we call the initial search radius. Each secondary detection found this way then forms a detection pair with the primary detection (see Fig.~\hyperref[fig:static]{\ref*{fig:static}a}). Thus, any detection can be part of multiple detection pairs.
			
			During a tracking run, we treat each pair as a singular unit with its own location in time \mbox{$t_{i, i+1} = (t_i+t_{i+1})/2$} and space \mbox{$\vec{p}_\text{pair} = (\vec{p}_{\text{det}, i}+\vec{p}_{\text{det}, i+1})/2$}, where $\vec{p}_{\text{det}, i}$ and $t_i$ are the positions and recording times of its two detections, \mbox{$i \in \{0, 2, 4, ..., N-2\}$} (for simplicity we refer to any time-step as $t_i$ following Eq.~\ref{eq:velocity}). What discriminates pairs from single detections however, is that we additionally attribute each pair with a velocity vector:
			
			\begin{equation}
				\label{eq:velocity} \vec{v}_{\text{pair}} = \frac{\vec{p}_{\text{det}, i+1} - \vec{p}_{\text{det}, i}}{t_{i+1} - t_i}.
			\end{equation} 
		
			Once all the pairs are created and their properties computed, they make up the initial pool of available pairs. Because pairs that stem from the same two images all share the same point in time, the pool of available pairs is quantized into $N/2$ pair groups (see Fig.~\ref{fig:data_reduction}). 
			
			Each pair from this pool is considered as part of a candidate track at least once: either to establish a new one, or to become part of another. We allow detections and pairs to be associated with only one track however, thus as soon as a candidate track is accepted, its components (and any other unrelated pairs its detections were part of) become unavailable throughout the rest of the tracking run. 
			
			\begin{figure*}[!ht]
	  			\centering
				\includegraphics[width=\linewidth]{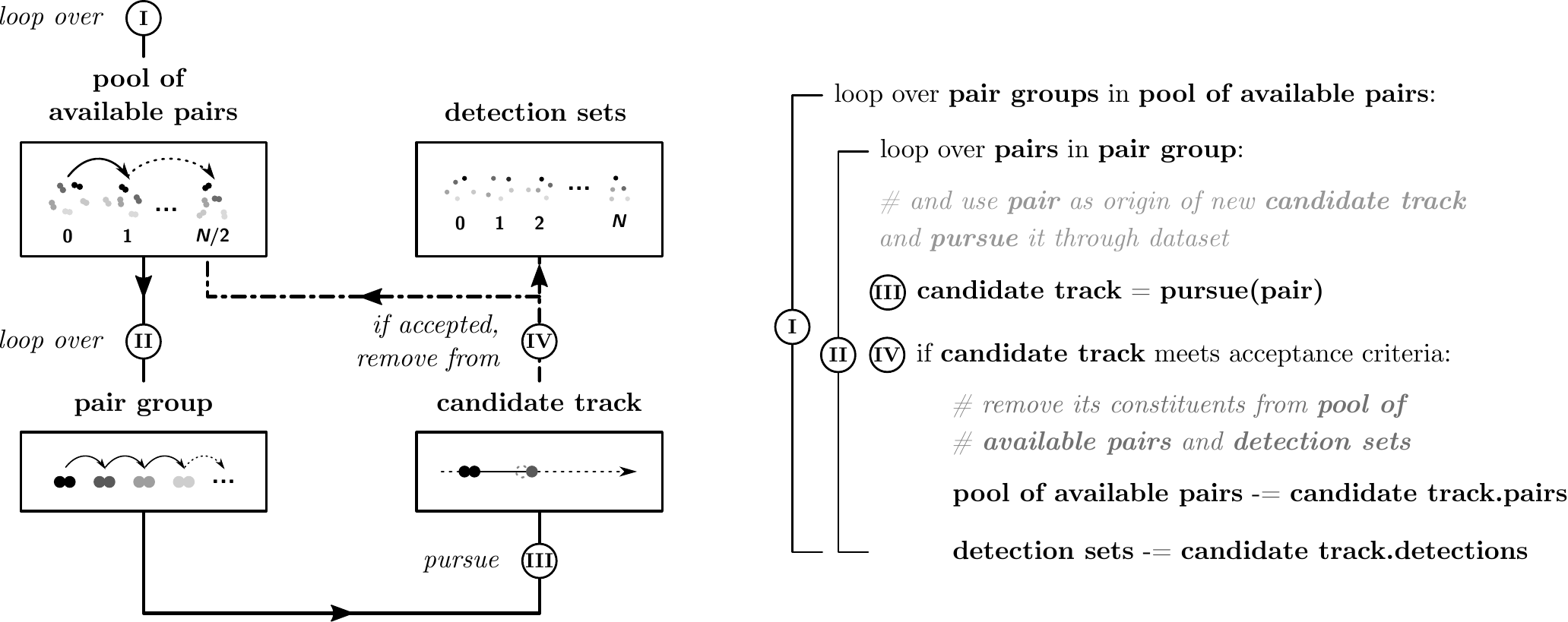}
				\caption{Flowchart and pseudo-code illustrating the structure of an entire tracking run. The algorithm iterates over the pool of available pairs (I), using each pair as the starting point for a new candidate track (II, III). If a candidate track is accepted (IV), its pairs and detections are removed from the respective sources (as well as any other available pair that shares detections with the track) and cannot be used to create future tracks.}
				\label{fig:tracking_run}
			\end{figure*}

			\begin{figure*}[!ht]
	  			\centering
				\includegraphics[width=\textwidth]{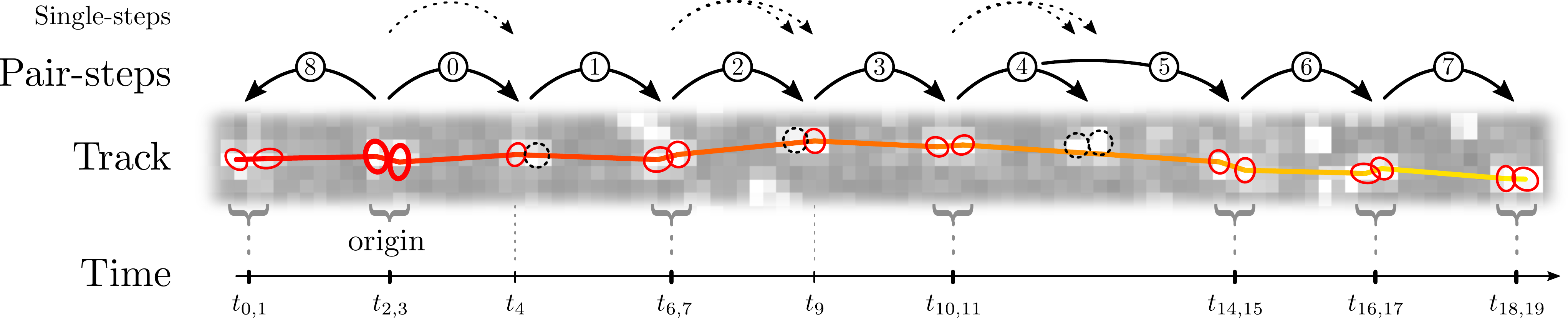}
				\caption{Typical track demonstrating the pursuit process. The algorithm operates pair-step-wise, going from one pair group to the next. In this case, it starts with a pair from the second group. Because the origin lies in the first half of the image sequence, the candidate track is first tracked forward, then backward in time (indicated by the circled numbers). If no suitable pair is found at a given step, the algorithm switches to searching for suitable single detections (single-step) instead, starting with the detection set that is closer in time to the previous step. The algorithm only searches the second set as well if it finds no suitable detection in the first. Afterwards, the pair-tracking continues. The track's color gradient from red to yellow indicates the direction of time (and therefore the object's motion through space). While the red ellipses mark the detections that make up the track, the black, dashed circles indicate where detections are missing. The background image is part of the stacked image similar to Fig.~\ref{fig:stack}.}
				\label{fig:pursuit}
			\end{figure*}
			
			Accordingly, a complete tracking run consists of a series of individual pursuits of one candidate track at a time (see Fig.~\ref{fig:tracking_run}). The algorithm walks forward in time through the pool of available pairs, and starts a new candidate track with each (though successful pursuits lower the number of remaining available pairs). We call this initial pair of a candidate track its origin, and track from it forward and backward in time.

			Each candidate track is pursued from one pair group to the next, an operation quantized in what we call pair-steps (see Fig.~\ref{fig:pursuit}). If at any pair-step no suitable pair could be found, the algorithm switches for that instance to single-steps, looking for a single suitable detection instead. Afterwards, the algorithm switches back to pair-tracking, searching the next pair group in line. 
			
			In this manner, each candidate track is pursued throughout the whole dataset, independent of how many pairs or detections may have been missed along the way. The pursuit only stops prematurely if, after no suitable pair or detection were found at a given step, it is determined that the center of the search area lies outside the dust field. The pursuit at the other end of the candidate track remains unaffected by this. Any pursuit concludes by checking if the candidate track qualifies as a track (see below). Only then does the algorithm move on to pursue a new candidate track.

		\subsection{Tracking parameters}

			Whether a candidate track qualifies as a track and which criteria single detections and pairs need to satisfy to become part of one is governed by a set of tracking parameters. While some of them are static and do not change during the whole tracking run, others are dynamic and adjust as the candidate track in pursuit evolves. The static tracking parameters only play a role at the beginning and end of a pursuit. They are (see Fig.~\ref{fig:static}): 
						 		
			\begin{description}
				\item[The initial search radius $R_\text{init}$,] which is used to create the detection pairs (Fig.~\hyperref[fig:static]{\ref*{fig:static}a}). It limits the maximum velocity any pair can have and sets the stage for individual pursuits of candidate tracks, as the properties of the origin are decisive in what the algorithm is looking for. 
				\item[The residual offset $R_\text{off}$,] which is the final tracking parameter that affects the candidate track itself (Fig.~\hyperref[fig:static]{\ref*{fig:static}b}). Once the pursuit of a candidate track is over, a final curve is fitted to its detections, and the distances $d_\text{off}$ between them ($\vec{p}_{\text{det}, i}$) and the locations where they should lie according to the fit ($\vec{p}_\text{fit} (t_i)$) are calculated. Any detection where $d_\text{off} > R_\text{off}$ is removed from the candidate track.
				\item[The] minimum number of detections $N_\text{det}$ and detection pairs $N_\text{pair}$, which define the acceptance thresholds (Fig.~\hyperref[fig:static]{\ref*{fig:static}c}). After the residual offsets have been checked, any candidate track must have at least that many detections and detection pairs to be accepted. 
			\end{description}

			Even though exceeding the acceptance thresholds does not guarantee that a group of detections all belong to the same object, it does increase our confidence in the tracking results. The more detections a candidate track contains, the less likely it is that they are unrelated (i.e., stem from different particles or sources). Thus, for the remainder of the tracking run, we treat any candidate track that passes these thresholds as a valid track. 	

\begin{figure}[!ht]
	  			\centering
				\includegraphics[width=\linewidth]{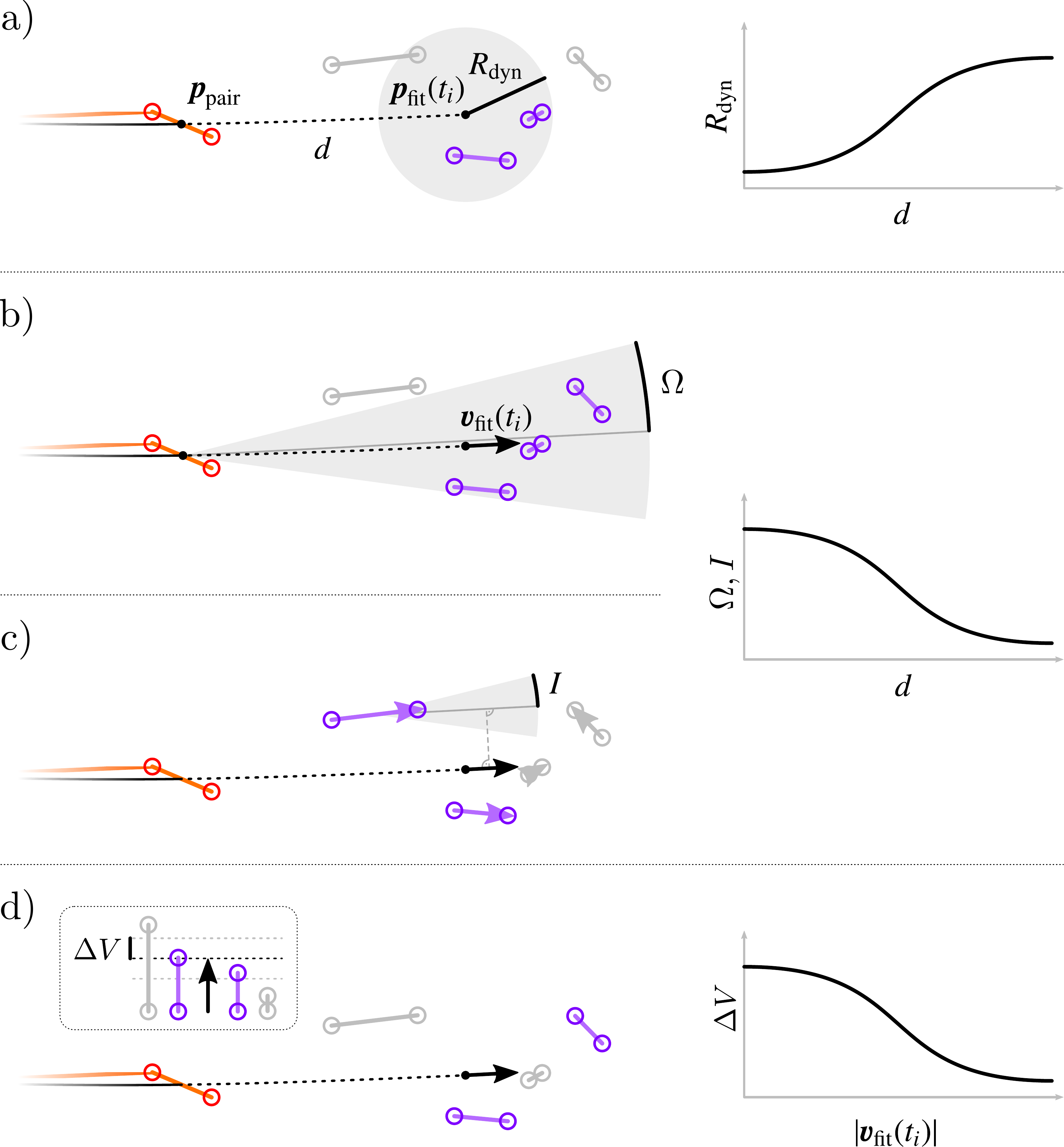}
				\caption{Diagrams illustrating how the dynamic tracking parameters operate (pairs that satisfy the respective criteria are shown in violet, the ones that do not in gray): a) the dynamic search radius $R_\text{dyn}$, which defines the area the algorithm searches for candidate pairs or detections. It depends on the distance $d$ between the candidate track's pair $\vec{p}_\text{pair}$ closest in time to the investigated step, and the predicted position $\vec{p}_\text{fit} (t_i)$ where the next pair is expected to lie according the curve (black, partly dashed line) fitted to the track (orange path). The relation between $R_\text{dyn}$ and $d$ is that of an arctangent (see Eq.~\ref{eq:dmax}) shown by the graph on the right. b) the (maximum) offset angle $\Omega$, which defines a circular sector within which candidate pairs or detections must lie. The sector originates from the candidate track's closest pair and opens up in the same direction as the fitted curve at the investigated time-step ($\vec{v}_\text{fit} (t_i)$, black arrow). The offset angle also depends on $d$ in the form of an arctangent, although reversed, as shown by the graph on the right (see Eq.~\ref{eq:amax}). c) the (maximum) inclination angle $I$, whose value is equal to that of $\Omega$. It defines the maximum inclination candidate pairs can have with respect to $\vec{v}_\text{fit} (t_i)$. d) the relative difference in speed $\Delta V$, which determines how much the speed of a candidate pair can relatively deviate from $|\vec{v}_\text{fit} (t_i)|$. The relation between $\Delta V$ and $|\vec{v}_\text{fit} (t_i)|$ is also that of a reversed arctangent as shown by the graph on the right (see Eq.~\ref{eq:vmax}).}
				\label{fig:dynamic}
			\end{figure}

			Avoiding to add unrelated detections is also helped by the dynamic tracking parameters: They repeatedly adjust to the properties of the candidate track during its pursuit and therefore narrow the track-specific parameter space that the algorithm searches for suitable detections and detection pairs. With the exception of the first pair-step, where the properties of the origin are used, these parameters depend on the properties of a curve that is fitted to the candidate track at every step. We refer to the detections or detection pairs that satisfy the criteria derived from these parameters as candidate pairs or detections. The dynamic tracking parameters are (see Fig.~\ref{fig:dynamic}):
			
			\begin{figure}[!ht]
	  			\centering
				\includegraphics[width=\linewidth]{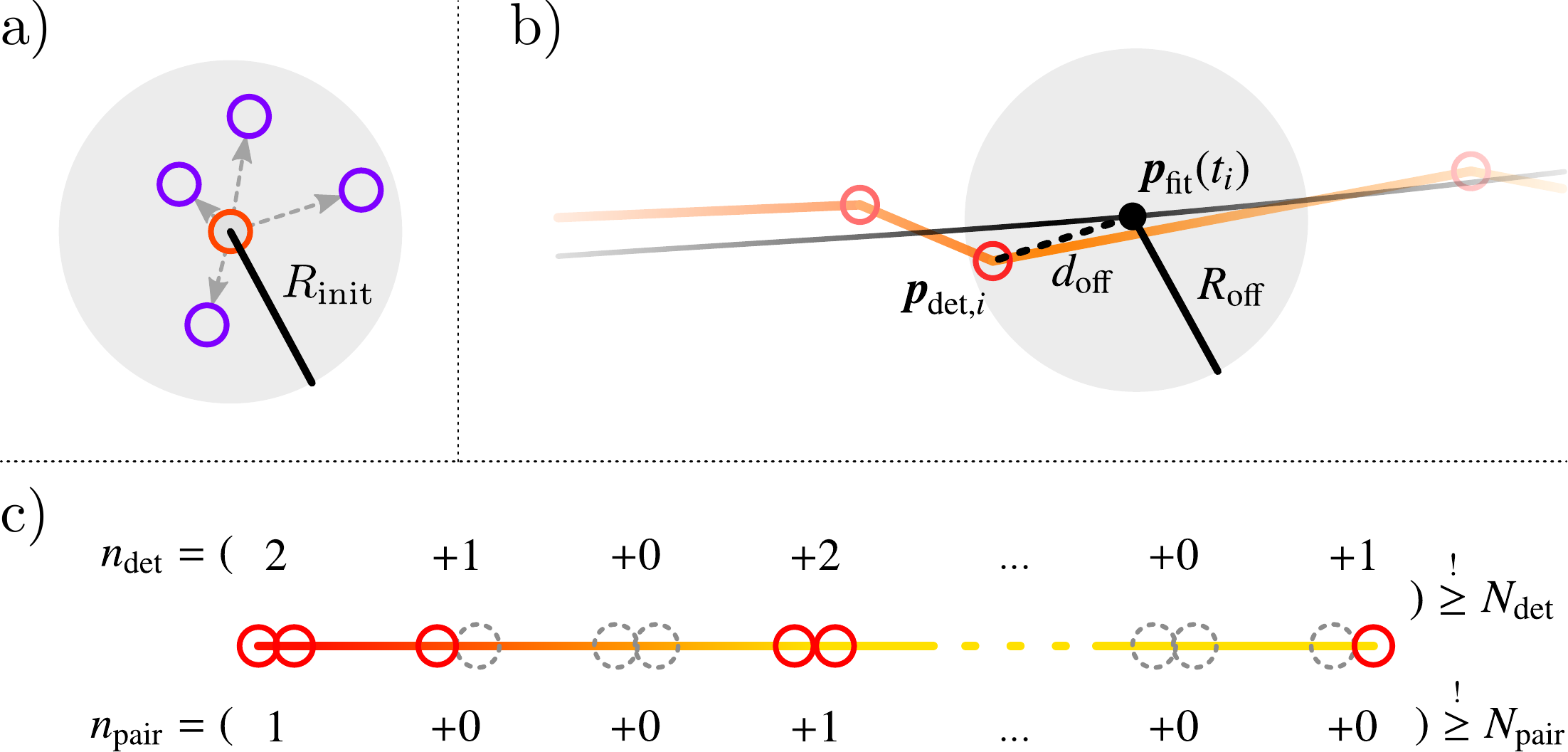}
				\caption{Diagrams illustrating how the static tracking parameters operate: a) the initial search radius $R_\text{init}$ around a primary detection (orange circle), used to create pairs with secondary detections (violet circles); b) the residual offset $R_\text{off}$, which defines the maximum distance $d_\text{off}$ any detection $\vec{p}_{\text{det}, i}$ (red circle) can have from the corresponding location $\vec{p}_\text{fit} (t_i)$ of the curve (black line) fitted to the candidate track (orange path); and c) the minimum number of detections $N_\text{det}$ and detection pairs $N_\text{pair}$, which any candidate track ($n_\text{det}$, $n_\text{pair}$) must have to be accepted as a track. The candidate track is shown as a gradient line from red to yellow, the present and missing detections as red, and gray dashed circles, respectively.}
				\label{fig:static}
			\end{figure}
			
			\begin{description}
				\item[The dynamic search radius $R_\text{dyn}$,] which defines the area within which the algorithm looks for candidate pairs and detections (Fig.~\hyperref[fig:dynamic]{\ref*{fig:dynamic}a}). The size of the area depends on the distance $d$: the spatial distance between the candidate track's pair $\vec{p}_\text{pair}$ that is closest in time to the investigated step (in case of single-tracking the closest detection), and the predicted position $\vec{p}_\text{fit} (t_i)$ of where the next pair (or detection) is expected to lie according to the curve fitted to the candidate track. We chose the relation between the search radius and distance to be that of an arctangent, whose free parameters $R_\text{min}$, $R_\text{max}$, $\hat{d}$ (shift), and $\bar{d}$ (stretch) we can control at the beginning of the tracking run:
				\begin{equation}
					\label{eq:dmax} R_\text{dyn}(d) := R_\text{min} + \left(\frac{R_\text{max} - R_\text{min}}{2}\right)\left(1 + \frac{2}{\pi}\arctan\left(\frac{d - \hat{d}}{\bar{d}}\right)\right).
				\end{equation}
				Increasing with distance, this function still allows for meaningful search radii at the smallest distances, while being capped at larger distances, as to not include candidate pairs or detections that are too far out.
                     
				\item[The (maximum) offset angle $\Omega$,] which defines a circular sector within which the candidate pairs or detections must lie (Fig.~\hyperref[fig:dynamic]{\ref*{fig:dynamic}b}). It measures from the vector that points in the same direction as the fitted curve at the time of the investigated step ($\vec{v}_\text{fit} (t_i)$), and, like the dynamic search radius, it depends on the distance $d$. The sector originates from the center of the candidate track's closest pair (or closest detection in case of single-tracking), and opens up in tracking direction. Its arc spans twice the offset angle. The relation between $\Omega$ and $d$ is otherwise identical to that of Eq.~\ref{eq:dmax}, but with the arctangent flipped:
				\begin{align}
					\label{eq:amax} \nonumber &\Omega(d) := \\ &\Omega_\text{min} + \left(\frac{\Omega_\text{max} - \Omega_\text{min}}{2}\right)\left(1 - \frac{2}{\pi}\arctan\left(\frac{d - \hat{d}_\Omega}{\bar{d}_\Omega}\right)\right),
				\end{align}
				where we again have control over the free parameters $\Omega_\text{min}$, $\Omega_\text{max}$, $\hat{d}_\Omega$, and $\bar{d}_\Omega$. In this case however, we allow the largest deviations for the smallest distances, since even small positional changes perpendicular to the candidate track can mean large angular ones. The opposite is true for large distances. 	
									
				\item[The (maximum) inclination angle $I$,] which is also measured with respect to the candidate track's direction at the investigated time-step ($\vec{v}_\text{fit} (t_i)$, Fig.~\hyperref[fig:dynamic]{\ref*{fig:dynamic}c}). It shares the same value as $\Omega$, but instead limits the inclination that pairs can have toward the reference vector. Because single detections do not have an inclination, this parameter is only relevant during pair-tracking.
									
				\item[The relative difference in speed $\Delta V$,] which restricts how much the speed of candidate pairs can stray from that of the fitted curve at the investigated time-step ($|\vec{v}_\text{fit} (t_i)|$, Fig.~\hyperref[fig:dynamic]{\ref*{fig:dynamic}d}). It is calculated as the relative deviation from $|\vec{v}_\text{fit} (t_i)|$ in percent, from a relation that has the same shape as Eq.~\ref{eq:amax}:
				\begin{align}
					\label{eq:vmax} \nonumber &\Delta V(|\vec{v}_\text{fit}|) := \\ &\Delta V_\text{min} + \left(\frac{\Delta V_\text{max} - \Delta V_\text{min}}{2}\right)\left(1 - \frac{2}{\pi}\arctan\left(\frac{|\vec{v}_\text{fit}| - \hat{v}}{\bar{v}}\right)\right),
				\end{align}					 
				where we also have control over the free parameters $\Delta V_\text{min}$, $\Delta V_\text{max}$, $\hat{v}$, and $\bar{v}$. Analogous to the offset and inclination angle, we allow the largest relative deviation for the smallest speeds, because in this regime, pixelization and uncertainties in the pointing of the camera and source detection can have a significant effect. For high speeds on the other hand, we only expect small deviations, for example due to a curved flight path. Because single detections cannot be assigned a velocity, this parameter is also only relevant during pair-tracking.
			\end{description}
			
			For each candidate pair or detection that satisfies all criteria set up by the dynamic tracking parameters, we compute a match-factor $M$ as a proxy for the candidate's validity. The match-factor is used in two ways: to decide between different candidate pairs or detections, and to weigh the contribution of the selected one on the curve fitted to the candidate track:
			\begin{equation}
				\label{eq:pair_match} M_\text{cand} := 1 - \frac{1}{4} \left(\frac{r_\text{cand}}{R_\text{dyn}} + \frac{\omega_\text{cand} + I_\text{cand}}{\Omega} + \frac{\Delta v_\text{cand}}{\Delta V}\right),
			\end{equation}
			for pair-tracking, or
			\begin{equation}
				\label{eq:single_match} M_\text{cand} := 1 - \frac{1}{2} \left(\frac{r_\text{cand}}{R_\text{dyn}} + \frac{\omega_\text{cand}}{\Omega}\right),
			\end{equation}	
			for single-tracking, where $r_\text{cand}$, $\omega_\text{cand}$, $I_\text{cand}$ and $\Delta v_\text{cand}$ are the dynamic parameter values of the candidate pair or detection, which are normalized by the respective maximum values as determined by Eqs.~\ref{eq:dmax}--\ref{eq:vmax}. We then choose the pair or detection with the highest match-factor to become part of the candidate track.

		\subsection{Principal and extended tracking}
        \label{sect:extended_tracking}
        
            To address the different time-steps of the two sub-sequences (see Fig.~\ref{fig:timeline}), our tracking algorithm has two operating modes on the pursuit level: principal, and extended tracking. Due to the shorter intervals of the principal sequence, particle tracks are generally easier to identify during principal tracking--both visually and by the tracking algorithm. Thus, candidate tracks are only pursued during extended tracking if they passed the acceptance thresholds after principal tracking. For the same reason, any track becomes part of the final tracking results independent of how many detections were missed during this second stage. 
            
            Both modes have their own set of predefined tracking parameters. Extended tracking however has an additional parameter we call life. The lives of a track define how many detection pairs are allowed to be missed during extended tracking. If no pair and no single detection is found at a given step, then the life counter is reduced by 1. Lives also cannot be replenished: should the counter fall to zero, the pursuit is stopped. This prevents adding unrelated detections to a track, something that becomes increasingly more likely the further the search area is away from the established part of the track.
            
            Once the algorithm checked the residual offset again after extended tracking, another control mechanism executes. Because detections may have been removed during the residual offset check, the extended part of the track is inspected for larger gaps (i.e., missing pairs). Should any individual gap or the sum of all gaps be larger than the granted extra lives, then all detections that come after the critical gap (i.e., the gap that let the sum of all gaps exceed the number of extra lives) are removed as well. 
            
            Finally, principal and extended tracking differ by the kind of curve that is fitted to the candidate track during its pursuit. Because the particles only travel relatively short distances during the principal sequence, we fit their tracks with straight lines. This is more robust than a parabola for example, since we found that the parabola's extra degree of freedom often causes the tracking algorithm to trail off in the wrong direction when the detections of a candidate track are not perfectly aligned. During the extended sequence however, we expect tracks to curve significantly because it covers a much longer time period. Thus at this stage, we fit parabolas to the tracks, which is also less likely to fail now that the tracks already consist of a considerable amount of detections. 
            
            The curve fitted to a candidate track to determine the residual offsets on the other hand is always a parabola. And no matter the circumstance under which a curve is fitted, the detections are always weighted by the match-factor (Eqs.~\ref{eq:pair_match} \& \ref{eq:single_match}) they were assigned when added to the track.

		\subsection{Sidereal-motion-based attitude correction}

			\begin{figure}[!b]
	  			\centering
				\includegraphics[width=\linewidth]{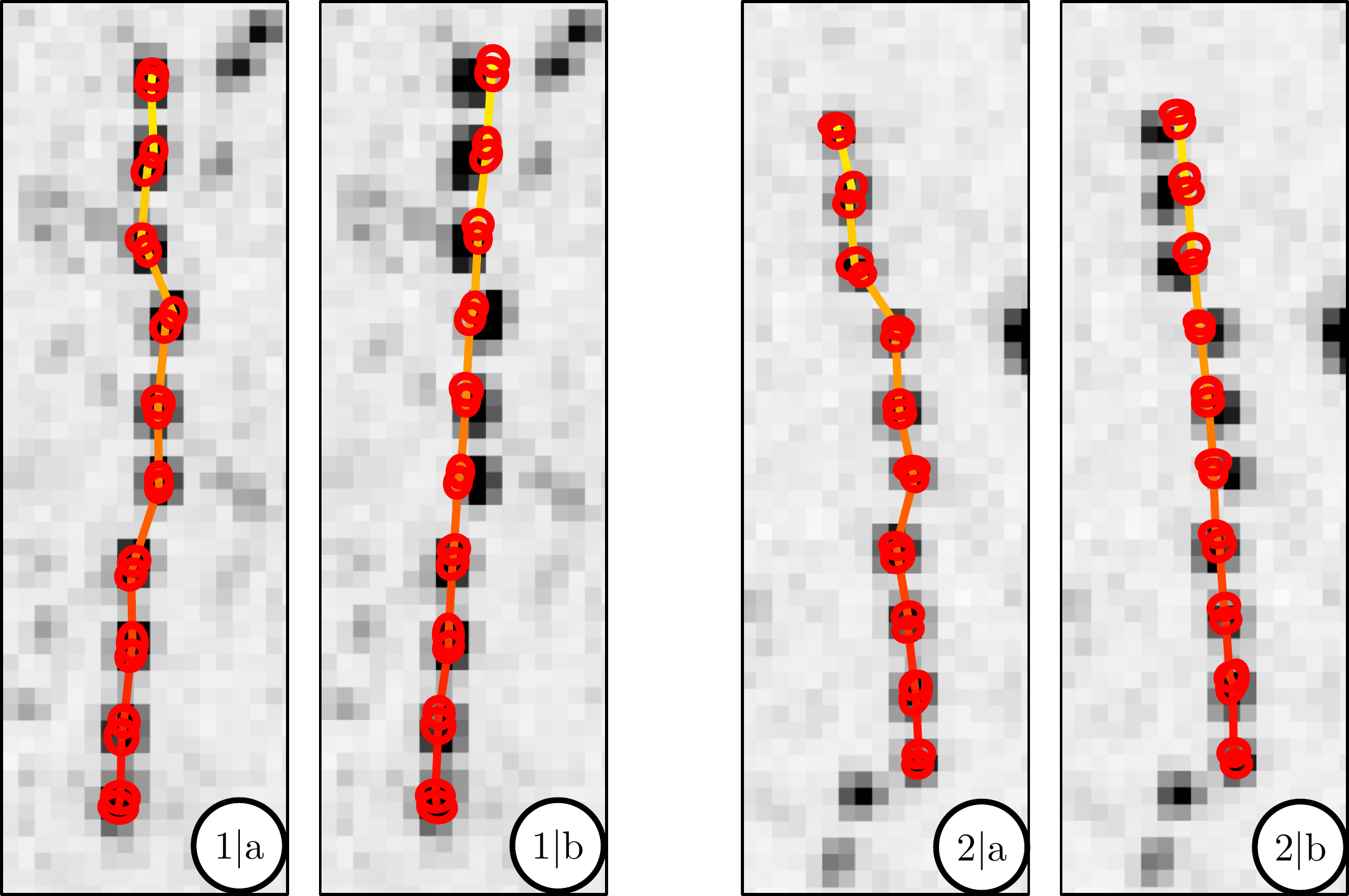}
				\caption{Two sample tracks (1, 2) from the principal sequence, once without (a) and once with (b) the pointing correction applied (the image shown in the background is left unchanged). Tracks are shown as colored lines from red to yellow, detections indicated by red ellipses.}
				\label{fig:correction}
			\end{figure}
			
		    This process precedes any particle tracking, but as it uses the same tracking algorithm, it is described only now. While studying the stacked image of sequence STP090, we noticed a pointing fluctuation with a typical amplitude of a few pixels that occurred during the principal sequence and which can be observed in every track (see Fig.~\ref{fig:correction}). It compromises the tracking results in several ways: (1) A significant spread of detections from their expected positions can quickly lead the algorithm to go off trail. (2) To account for a higher variance in location, velocity, and orientation of candidate pairs and detections, we need to chose more lenient tracking parameters. Inevitably, this further increases the chances of adding unrelated detections and going off trail. (3) Drastic changes in velocity also translate into incorrectly computed accelerations. This makes extended tracking based on parabolas virtually impossible, as predictions over the long time intervals between image pairs require accurate accelerations; otherwise, the search areas are too far off, again leading the candidate track to go astray.
            
			\begin{figure*}[!ht]
	  			\centering
				\includegraphics[width=\linewidth]{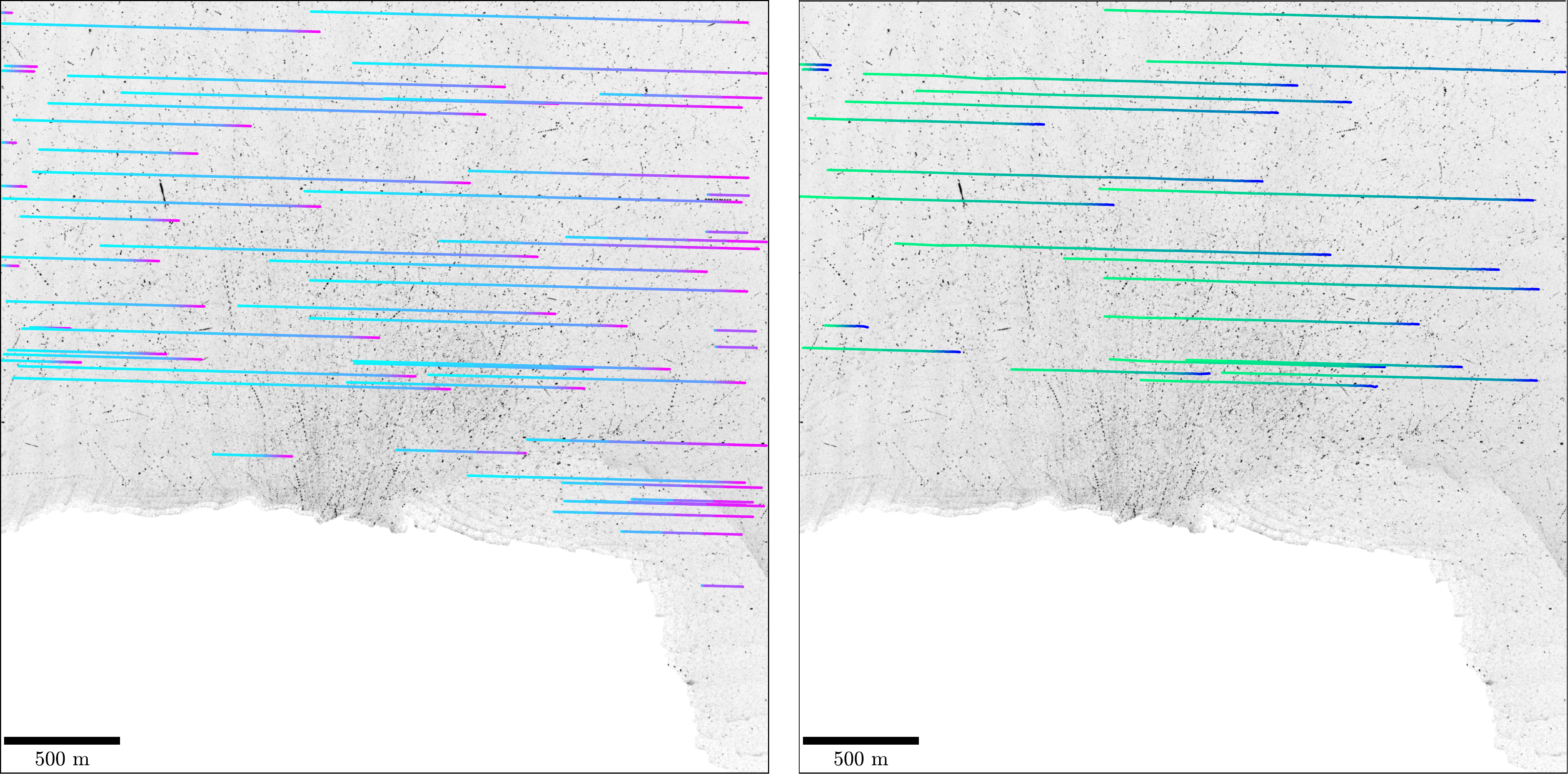}
				\caption{Sidereal objects identified in the dust field of sequence STP090. On the left: results from searching the SIMBAD database (lines colored violet to aquamarine). On the right: sidereal tracks obtained from our tracking algorithm that matched some of those results (lines colored dark blue to green). The objects move from right to left.}
				\label{fig:stars}
			\end{figure*}
			
			\begin{figure}[!ht]
	  			\centering
				\includegraphics[width=\linewidth]{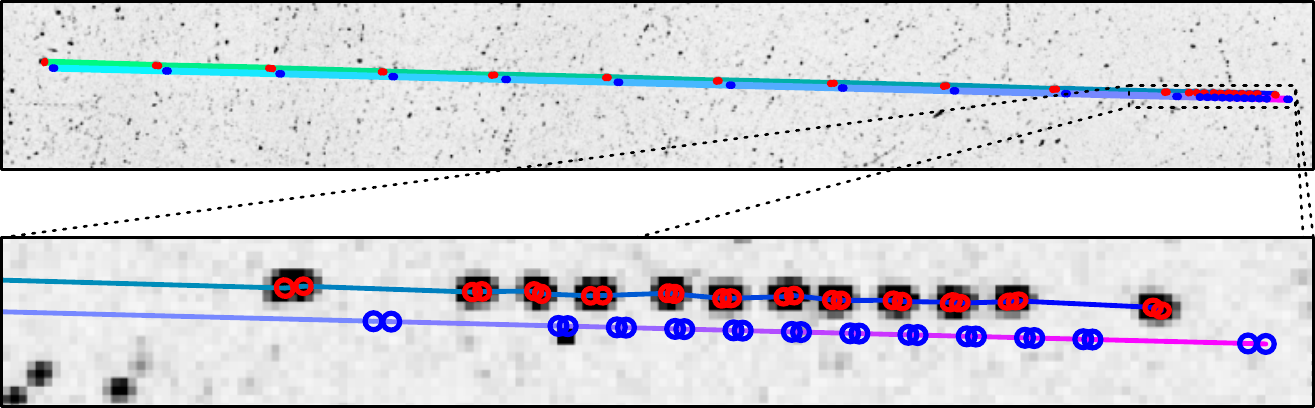}
				\caption{Sample of a sidereal track (top line from dark blue to green, detections indicated by red ellipses) and the expected motion of a sidereal object it was matched with (bottom line from violet to aquamarine, expected positions indicated by blue circles). The top panel shows the whole track, the bottom one a close-up of the first 24 detections, including the entire principal sequence. The pointing fluctuation is mainly acting along the direction of motion from right to left.}
				\label{fig:star_track}
			\end{figure}
		    			
			\begin{figure*}[!ht]
	  			\centering
				\includegraphics[width=\linewidth]{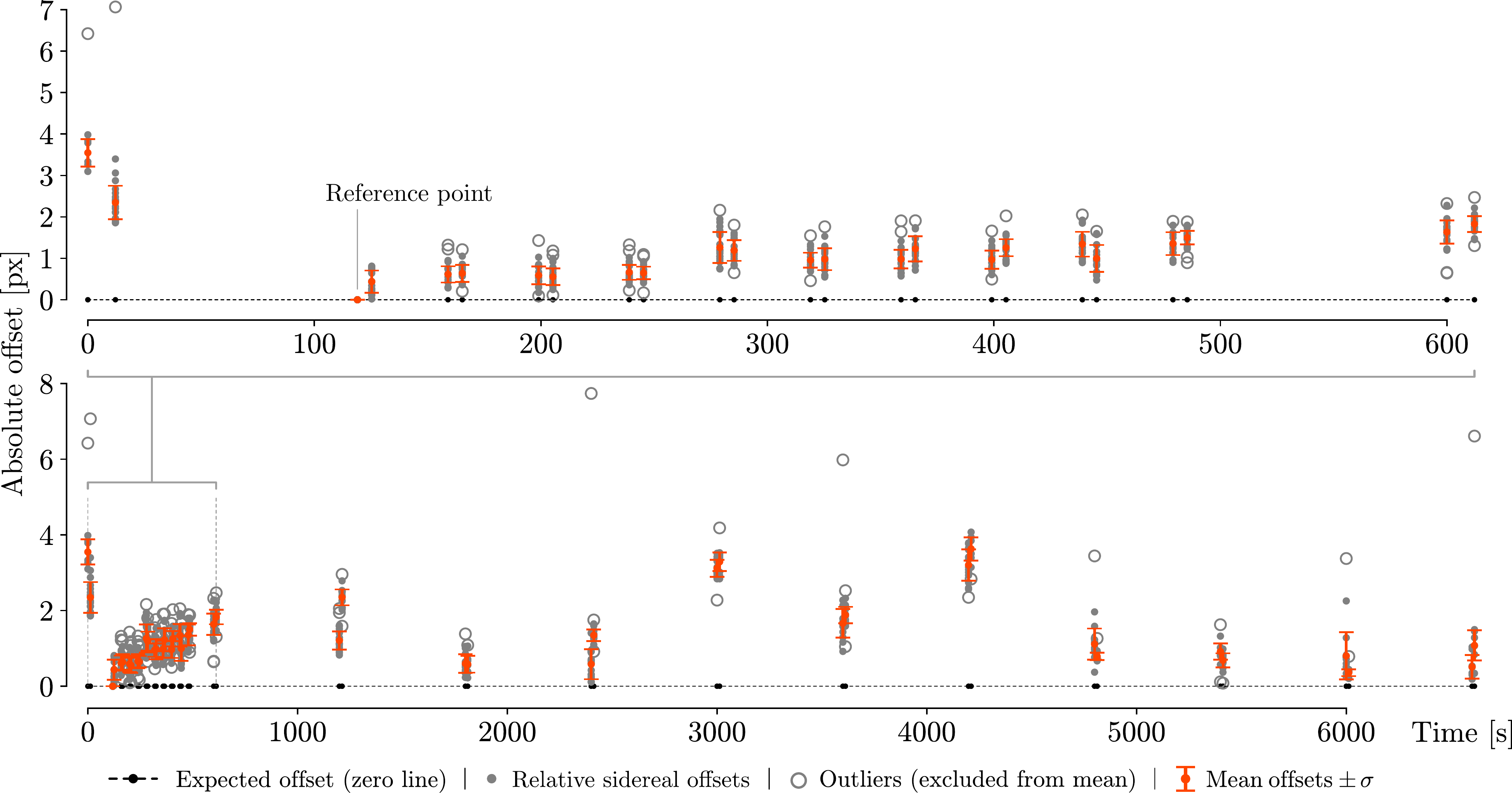}
				\caption{Pointing fluctuation derived from the apparent motion of 21 sidereal objects in the dust field of sequence STP090 (see Eqs.~\ref{eq:wobble1}--\ref{eq:wobble3}). The reference point indicates which image and therefore which detections we used to calculate the relative distances. The black circles and dashed line mark the zero line (no pointing fluctuation). The filled and the open gray circles show the measured offsets from sidereal tracks: while the filled ones were used to calculate the mean values that represent the pointing fluctuation (orange circles with errorbars), the open ones are outliers that were excluded from the calculations, as they are assumed to result from unrelated detections.}
				\label{fig:wobble}
			\end{figure*}
		    
		    Since the deviations are systematic, we attribute them to unexpected changes in spacecraft attitude. During STP090 (and other sequences like it), Rosetta's orbit around \object{67P} is noticeable. But to keep the camera's reference frame fixed to the comet's center of mass, the spacecraft's attitude was constantly adjusted. Sidereal objects therefore describe an apparent linear motion across the dust field. By comparing the pointing data of the image headers--which represent the commanded pointing--with the actual motion derived from tracks, we can thus reconstruct the pointing fluctuation.

		    To identify sidereal objects in a sequence, we query the SIMBAD Astronomical Database \citep{2000A&AS..143....9W} via the astroquery library \citep{2019AJ....157...98G}. Objects such as binaries that are spaced too close to each other to be distinguishable in the images are recorded only once. We then use gnomonic projection to transform the objects' equatorial coordinates back to image coordinates and generate the expected motions (see Fig.~\ref{fig:stars}). 
			
			Next, we run the tracking algorithm in a local area around each of the identified objects, visually compare the tracking results to the expected motions, and match them manually. Figure~\ref{fig:star_track} shows an example of such a track we call sidereal track, and its companion, the previously estimated motion. By choosing a reference image, calculating the relative distances of the commanded positions to that reference image, doing the same for the detections of the sidereal tracks, and subtracting the former distances from the latter, we can calculate the relative offsets induced by the pointing fluctuation:
			\begin{align}
			    \label{eq:wobble1} \vec{\delta}_i &= (\vec{p}_{\text{det}, i} - \vec{p}_{\text{det, ref}}) - (\vec{p}_{\text{com}, i} - \vec{p}_{\text{com, ref}}),
			\end{align}
			where $\vec{\delta}_i$ is the relative offset of a sidereal track at the $i$th image, $\vec{p}_{\text{det, ref}}$ the position of the track's detection in the reference image, and $\vec{p}_{\text{com}, i}$ the commanded position of the $i$th image relative to the position of the reference image $\vec{p}_{\text{com, ref}}$. 
			
			However, because we need to choose particularly liberal tracking parameters during the pursuit of sidereal tracks to account for the still present pointing fluctuation, there is an increased chance to pick up unrelated detections. Thus before we estimate the pointing fluctuation, we calculate the mean absolute offset values:
			\begin{align}
			    \label{eq:wobble2} \bar{|\vec{\delta}_i|} = \frac{1}{M_i} \sum_{j=0}^{M_i} |\vec{\delta}_{i, j}|,
			\end{align}
			where $M_i$ is the number of sidereal tracks that have a detection in the $i$th image, and exclude any data point from the signal estimation whose absolute offset lies outside a certain range around its mean ($\pm 1.7\,\sigma_i$ in case of sequence STP090). Only then do we calculate the mean offsets (see Fig.~\ref{fig:wobble}) and use them to correct our detection sets: 
			\begin{align}
			    \label{eq:wobble3} \bar{\vec{\delta}}_i = \frac{1}{\Tilde{M}_i} \sum_{j=0}^{\Tilde{M}_i} \vec{\delta}_{i, j},
			\end{align}
			where $\Tilde{M}_i$ is $M_i$ minus the excluded data points. Figure~\ref{fig:correction} shows two sample tracks from the principal sequence with and without the pointing correction.
			
			The choice of the reference point that is used to calculate the relative distances is crucial in this, since a) sidereal tracks that are missing the respective detection cannot be considered for the signature estimation, and b) sidereal tracks that have an unrelated detection as the reference end up with shifted offsets. For sequence STP090, we decided to use the first image of the principal sequence as the reference, as we found that detections from this image are usually not only included in all sidereal tracks but also the most reliable. 
		    
		    Finally, identifying the sidereal tracks also allows us to remove their constituents from the detection sets prior to the actual tracking run, ridding the tracking results of a statistical bias.

	\section{Parameter optimization}
	\label{sec:optimization}
	
	    The whole tracking process--including image cleaning, point source detection, attitude correction and the tracking run itself--involves far too many parameters ($>74$) for a systematic grid search. However for most parameters, preliminary tests indicate that their exact value is (within some range) secondary to achieving good results. We therefore focused on optimizing only the detection threshold (see Sect.~\ref{sec:detection}) and 15 dynamic tracking parameters of the principal and extended tracking (in the following referred to as principal and extended parameters, cf. Table~\ref{tab:parameters}), which we found to be more influential. In the following, we analyze their effect using a single quality index: the miss-rate $\Gamma$. It measures the percentage of detections that were missed during the pursuit of a track (i.e., whenever no suitable pair or detection was found, or when detections were later-on removed during offset checks):
		\begin{align}
		    \label{eq:miss_rate} \Gamma := 100 \cdot \frac{\tilde{N}_\text{det} - n_\text{det}}{\tilde{N}_\text{det}},
		\end{align}
		where $\tilde{N}_\text{det} \leq N$ is the maximum possible number of detections a specific track can have, which depends on whether and when the track supposedly left the dust field (e.g., if it lies close to the edge, the detections expected outside the field do not count toward the total). 
		We then estimated the quality of tracking results by looking not only at the total number of tracks, but more importantly at the numbers of tracks with \mbox{$\Gamma=0\,\%$} and \mbox{$\Gamma<30\,\%$}. By visual inspection we found 30\,\% to be a reasonable threshold where most tracks still belong to real particles and only occasionally incorporate unrelated detections.
	    
	    Because the principal parameters directly affect the total number of tracks (as the acceptance criteria are applied only once after the principal tracking), we optimized them first. Each of the twelve free parameters from Eqs.~\ref{eq:dmax}--\ref{eq:vmax} was varied at least ten times around an initial guess. Since testing all value combinations would still take $10^{12}$ individual tracking runs, we decided on a different strategy: First, we reduced the tracking runs to principal tracking only; and second, we tested each parameter value only once, keeping all other parameters constant. After the full value range for a given parameter was explored, we chose the value that produced the best results and used it as the parameter's new fixed value for the remaining runs. The results of this process are listed in Table~\ref{tab:parameters} in the appendix.
        
        Next, we optimized the extended parameters. Since we decoupled principal and extended tracking, and on its own the latter runs much faster than the former, we adapted our approach. Instead of testing the whole set, we only varied the three parameters that we deemed the most influential ($R_\text{max}$, $\Omega_\text{max}$, and $\Delta V_\text{max}$), and used the optimized principal values for the rest. We again chose ten different values for each of the three variable parameters, but this time, we explored all of the 1000 corresponding value combinations. The parameter set that produced the best results according to our miss-rate criteria is also shown in Table~\ref{tab:parameters}.
        
        Lastly, we estimated the optimal detection threshold (in units of signal-to-noise $S/N$). This dimensionless parameter determines the sensitivity of the detection algorithm toward weaker sources and is therefore directly linked to the number of detections per image. While being able to detect weaker sources can be beneficial in case of fainter particles or oblate rotators (i.e., particles that strongly vary in brightness), it also means to pick up more noise. Hence the detection threshold can neither be too high, as a significant portion of signal would be ignored, nor too low, as the signal would be overwhelmed by noise.
        
        So to optimize the detection threshold, we again chose ten values around an initial guess and measured the average detection density (within the dust field). The detection density however also strongly depends on exposure time ($T_\text{exp}$). In case of sequence STP090, the average detection density for images with \mbox{$T_\text{exp}=6$}\,s was roughly twice of that for images with \mbox{$T_\text{exp}=0.24$}\,s. Thus to keep the detection densities roughly constant, we adopted two separate detection thresholds, one for each exposure time. The one for \mbox{$T_\text{exp}=6$}\,s was then adjusted so that its corresponding detection densities would approximately match those of the \mbox{$T_\text{exp}=0.24$}\,s one. The detection sets produced by each of the ten threshold pairs then underwent their own attitude correction before their tracking runs were started (using the parameters listed in Table~\ref{tab:parameters}). As with the previous optimization processes, we surveyed the total number of tracks, and the numbers of tracks with \mbox{$\Gamma=0\,\%$} and \mbox{$\Gamma<30\,\%$}. After inspecting the most promising results more closely, we found that the best were produced by detection thresholds of $S/N=2.7$ (\mbox{$T_\text{exp}=6$}\,s) and \mbox{$S/N=3.6$} (\mbox{$T_\text{exp}=0.24$}\,s). They roughly correspond to an average detection density of $27.12 \times 10^{-4}$ detections per pixel, or about 7000 detections per image. 
        
\begin{figure}[!b]
  			\centering
			\includegraphics[width=\linewidth]{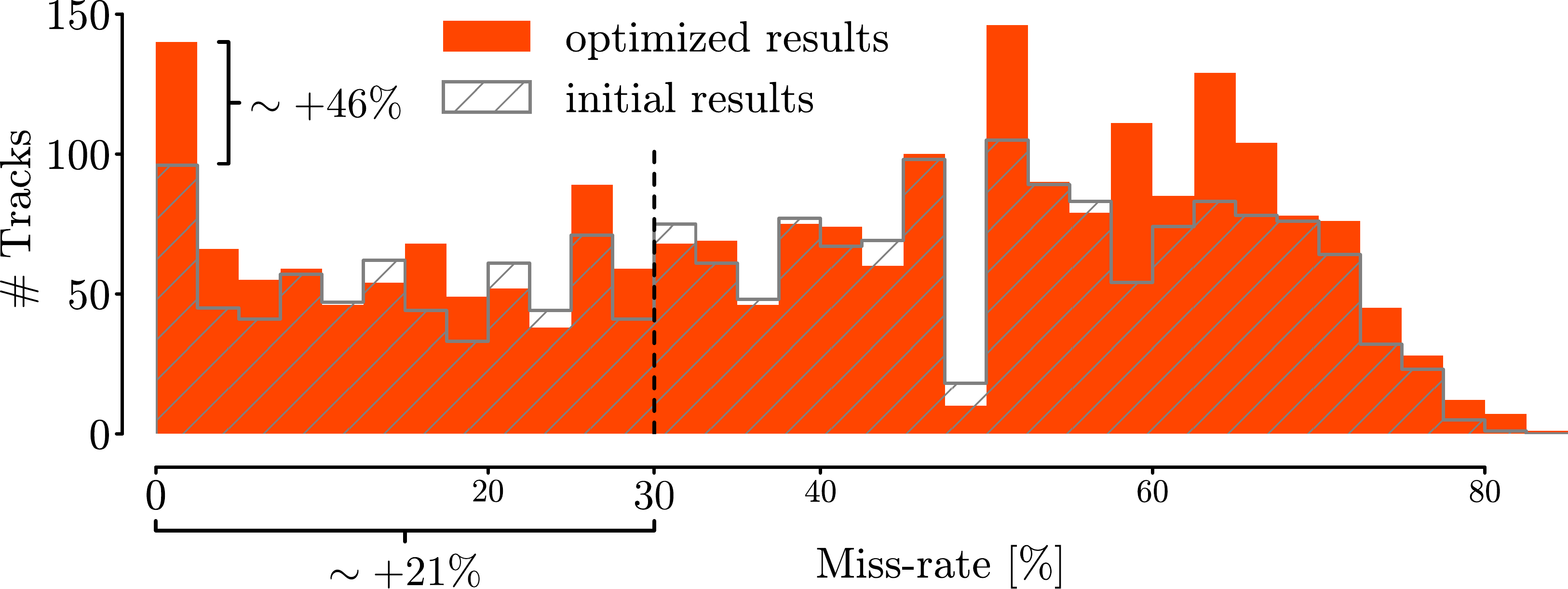}
			\caption{Effect of parameter optimization on the miss-rate distribution. Gray, hatched bars show the results produced by the initial set of tracking parameters, orange ones the optimized set.}
			\label{fig:optimization}
		\end{figure}

        Compared to the results produced by our initial set of tracking parameters (see Fig.~\ref{fig:optimization}), the optimization increased the total number of tracks by $\sim$\,18\,\% (from 1922 to 2268), the number tracks with \mbox{$\Gamma<30\,\%$} by $\sim$\,21\,\% (from 642 to 775), and the number of tracks without missing detections by $\sim$\,46\,\% (from 96 to 140)\footnote{While Fig.~\ref{fig:optimization} and the numbers discussed here are based on data produced by the latest version of the tracking algorithm, the optimization was unfortunately run on a previous version where the pointing fluctuation was slightly miscalculated due to a bug. However, because the error in the pointing fluctuation was small (\mbox{$<0.5$}\,px), and because the selected parameter values are only estimates of the optimal values that also work well with the correct pointing fluctuation, we decided against rerunning the optimization procedure.}.
        
	\section{Results and discussion} 
	\label{sec:results}

    	The goal of this study was to develop a robust algorithm to track dust particles of \object{67P} in image sequences recorded by OSIRIS NAC. As proof of concept, we applied the algorithm to sequence STP090 and optimized the tracking parameters. In the following, we first assess the general reliability of the tracking algorithm, and then give examples of how the tracking results can be evaluated to answer scientific questions.

    	\subsection{Algorithm assessment}

    		\subsubsection{Simulation}
    	    
        		\begin{figure}[!t]
          			\centering
        			\includegraphics[width=\linewidth]{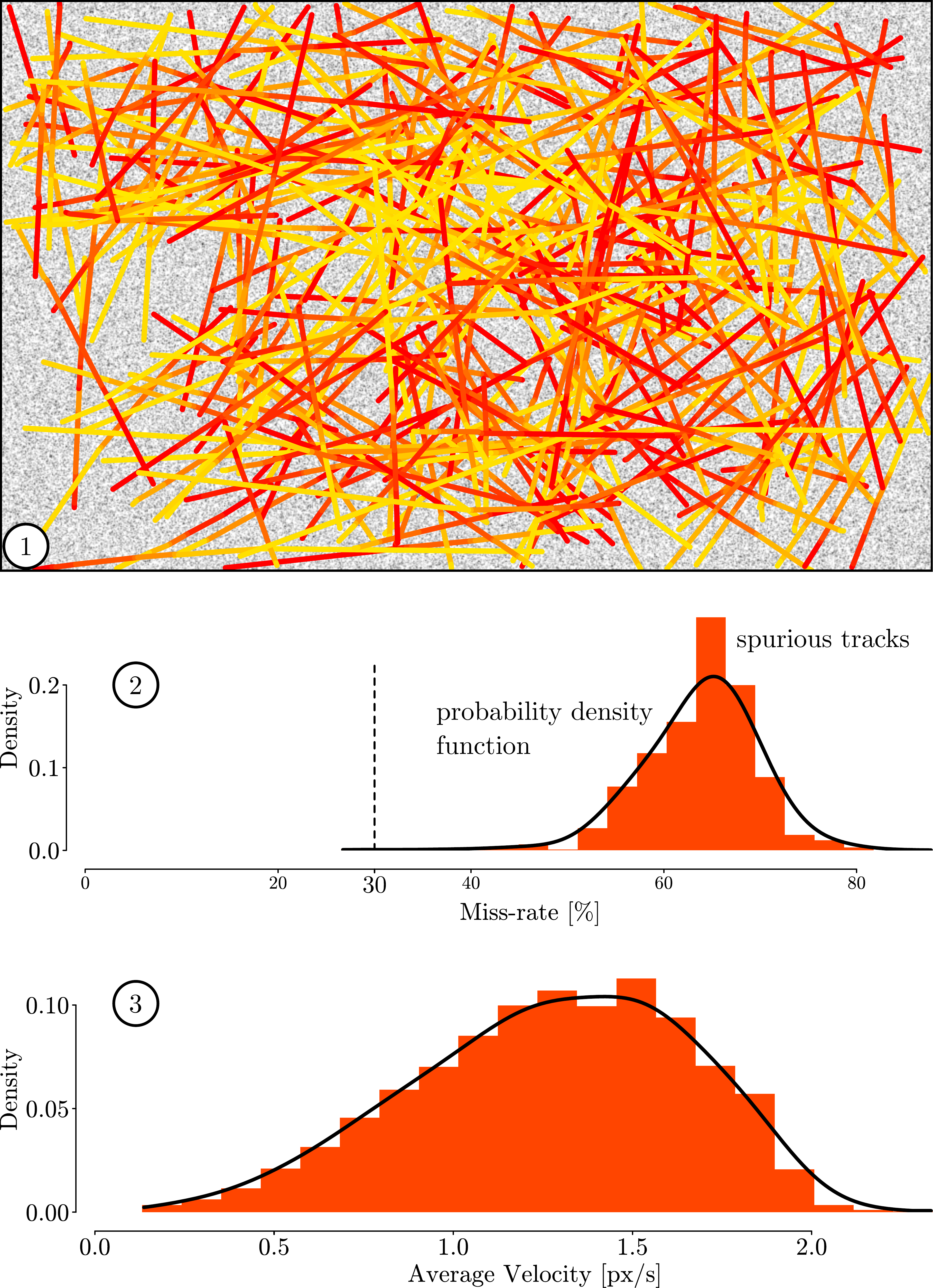}
        			\caption{Results from simulated data: (1) one of the ten simulations with a detection density of $27.59 \times 10^{-4}$\,det./px. The identified tracks are shown as gradient lines from red to yellow. The background image serves merely as a visual aid, showing the rough locations of detections (it is not a stack of images created to run the detection algorithm on; instead the detections were simulated first and the background image was created retroactively). (2) the combined miss-rate distribution from the ten simulations with a detection density of $27.59 \times 10^{-4}$\,det./px. (3) the velocity distribution of the same track population, showing a clear tendency of the algorithm to create more fast spurious tracks, especially when compared to velocity distributions from real data (Fig.~\hyperref[fig:track_stats]{\ref*{fig:track_stats}.2}). The probability density functions were created with Gaussian kernel density estimation.}
        			\label{fig:simulation}
        		\end{figure}
    	    
        	    To test our algorithm's tendency to create spurious tracks we simulated datasets which consisted entirely of random noise (i.e., "detections"), with detection densities ranging from $27.59 \times 10^{-4}$ to $39.42 \times 10^{-4}$\,det./px. Although the algorithm identified a few hundred to more than two thousand spurious tracks in the simulations depending on their detection density, it found few to none in the critical miss-rate regime below 30\,\%. In particular, we simulated ten different datasets for the detection density closest to that of the optimal detection thresholds ($27.59 \times 10^{-4}$\,det./px, Fig.~\ref{fig:simulation}). In those cases, only 260 tracks were found on average, and in total only 6 with \mbox{$\Gamma<30\,\%$}. 
    	    
    	    \subsubsection{Manual assessment}
    			
    			\begin{figure}[!t]
    	  			\centering
    				\includegraphics[width=\linewidth]{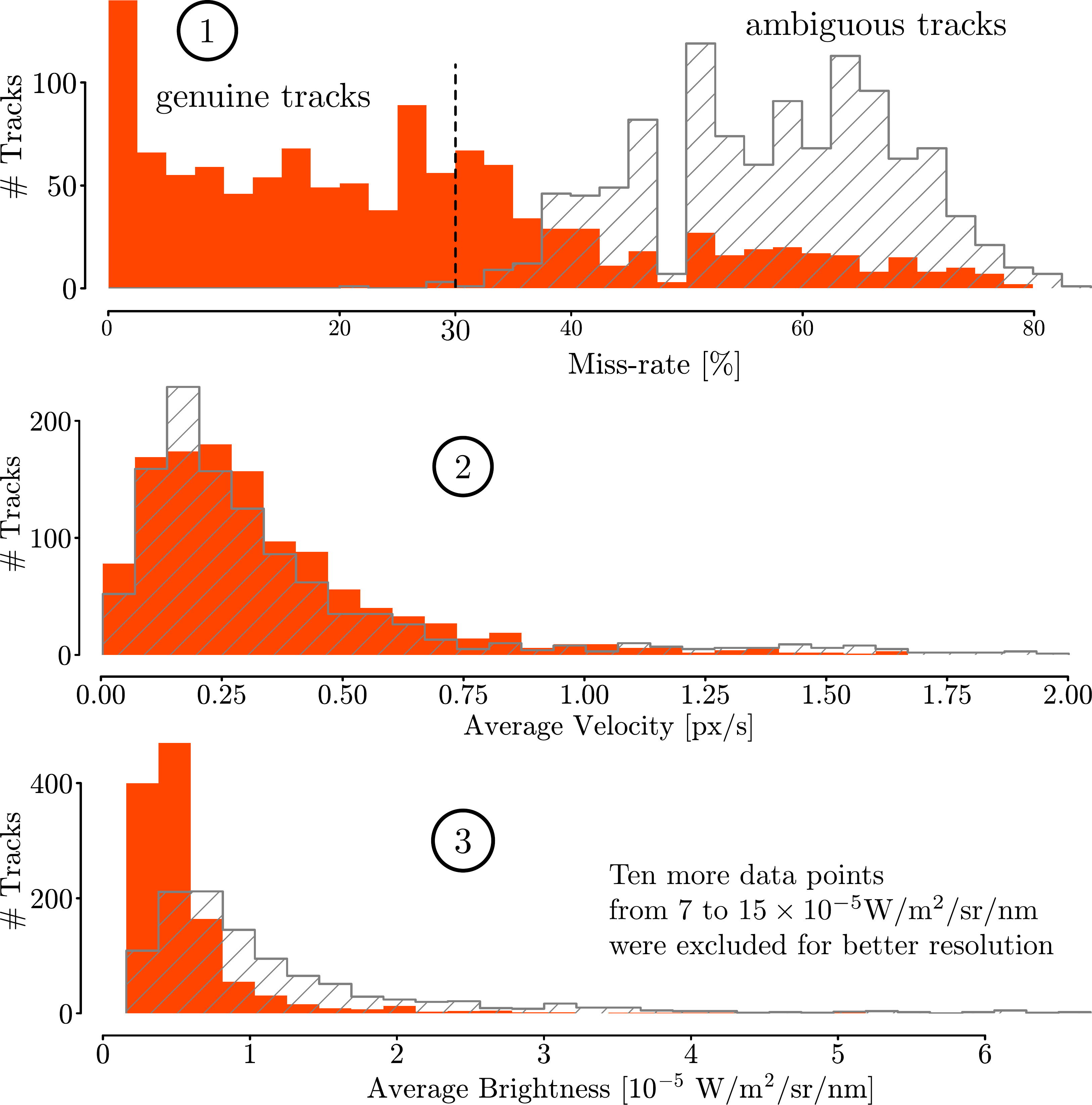}
    				\caption{Miss-rate (1), velocity (2), and brightness (3) distributions of tracks identified as either genuine (orange) or ambiguous (gray hatched).}
    				\label{fig:track_stats}
    			\end{figure}
    	    
        	    To further assess the reliability of our algorithm, we inspected and manually flagged each track found in sequence STP090 according to the following system: if they are (a) genuine, that is, whether we believe that they belong to actual particles and contain few to no unrelated detections, or if they are (b) ambiguous, that is, whether we believe that (the majority of) their detections do not belong to the same particle, stem from noise, or when it is impossible to tell.
            
        	    Of the 2268 tracks, we flagged 1081 ($\sim$\,48\,\%) as ambiguous, leaving 1187 ($\sim$\,52\,\%) as genuine. Figure~\hyperref[fig:track_stats]{\ref*{fig:track_stats}.1} shows that the miss-rate distributions of the ambiguous and genuine tracks have distinct shapes. In particular, only very few (4) of the ambiguous tracks have miss-rates less than 30\,\%. This is a good sign that our decision to base the parameter optimization on the number of tracks with \mbox{$\Gamma<30\,\%$} was appropriate. This is also further supported by the fact that the miss-rate distribution of the spurious tracks (Fig.~\hyperref[fig:simulation]{\ref*{fig:simulation}.2}) is very similar in shape to that of the ambiguous ones.
        	    
                Because manually judging the validity of tracks becomes increasingly difficult with the spread of their detections, we expect a bias against faster particles in the flagged tracks. Figure~\hyperref[fig:track_stats]{\ref*{fig:track_stats}.2} shows that such a trend seems to exist in our data, though only slightly. Figure~\hyperref[fig:simulation]{\ref*{fig:simulation}.3} on the other hand shows that our algorithm tends to create more fast than slow spurious tracks. Both effects probably contribute to the excess of fast ambiguous tracks.
                
                We also expect a bias toward flagging faint tracks more often as ambiguous. Figure~\hyperref[fig:track_stats]{\ref*{fig:track_stats}.3} however shows that the opposite was the case. This is likely caused by the overabundance of detections in the bright active area in the center of sequence STP090 (e.g., see Fig.~\ref{fig:source}): while of the genuine tracks only $\sim$\,15\,\% originate from here, of the ambiguous ones it is $\sim$\,22\,\% (the section indicated in Fig.~\hyperref[fig:selection_stats]{\ref*{fig:selection_stats}.1} was used to calculate those numbers).

    	\subsection{First results}
    	
    	    In the following, we present examples of how our tracking results can be used and interpreted. Since they mainly serve as a technical demonstration, we do not perform detailed analyses. Nevertheless, because \mbox{$\Gamma<30\,\%$} proved to be a good criterion to identify genuine tracks, we only consider the 775 tracks that satisfy it--more than three times as many tracks than were identified by \cite{10.1093/mnras/stw2179}.
    	
			\begin{figure}[!t]
	  			\centering
				\includegraphics[width=\linewidth]{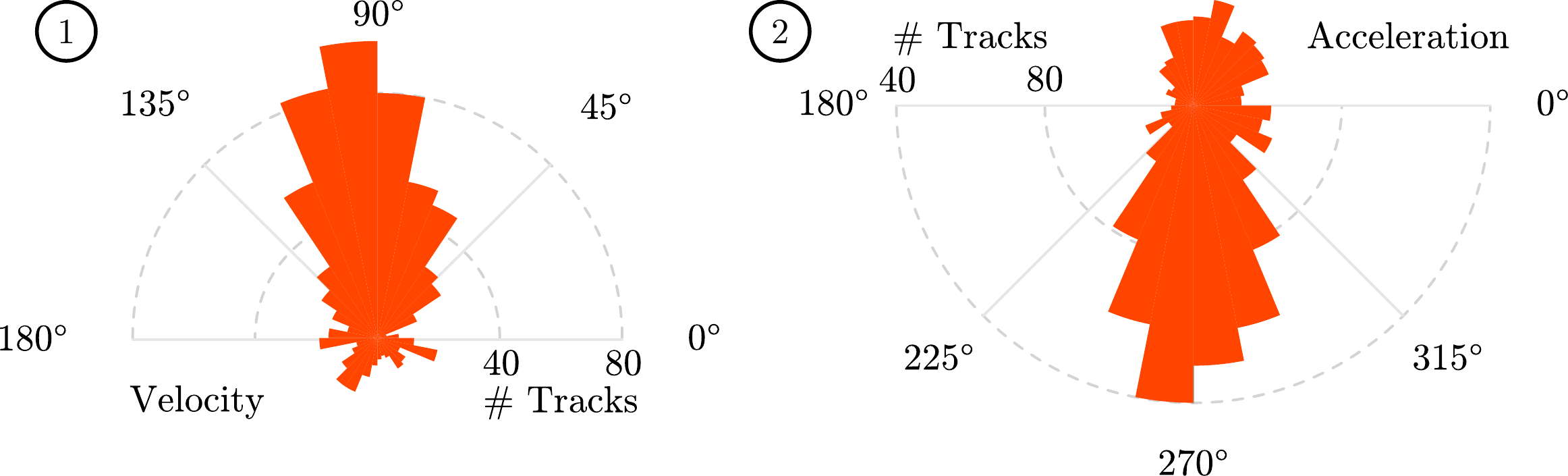}
				\caption{Angle distributions of the projected velocity (1) and acceleration (2) of all 775 selected tracks. The orientations of the diagrams coincide with how the images of sequence STP090 are displayed (i.e., $0\degree$ corresponds to the right direction, $90\degree$ to the up direction, etc.). While the projected velocity of most tracks seems to be pointing away from the nucleus, the acceleration of a similar number of tracks seems to be pointing toward it.}
				\label{fig:angles}
			\end{figure}
    	    
    	    Figure~\ref{fig:angles} shows the velocity- and acceleration-angle distributions of all 775 tracks. The projected velocity components of most tracks point upward, seemingly away from the nucleus and the central active area. This aligns well with what would be expected and \cite{10.1093/mnras/stw2179}'s findings. The projected acceleration components on the other hand mostly point downward and seem to be dominated by the nucleus' gravity.
    	    
    	    \begin{figure}[p]
                \centering
                \subfloat{\includegraphics[width=\linewidth]{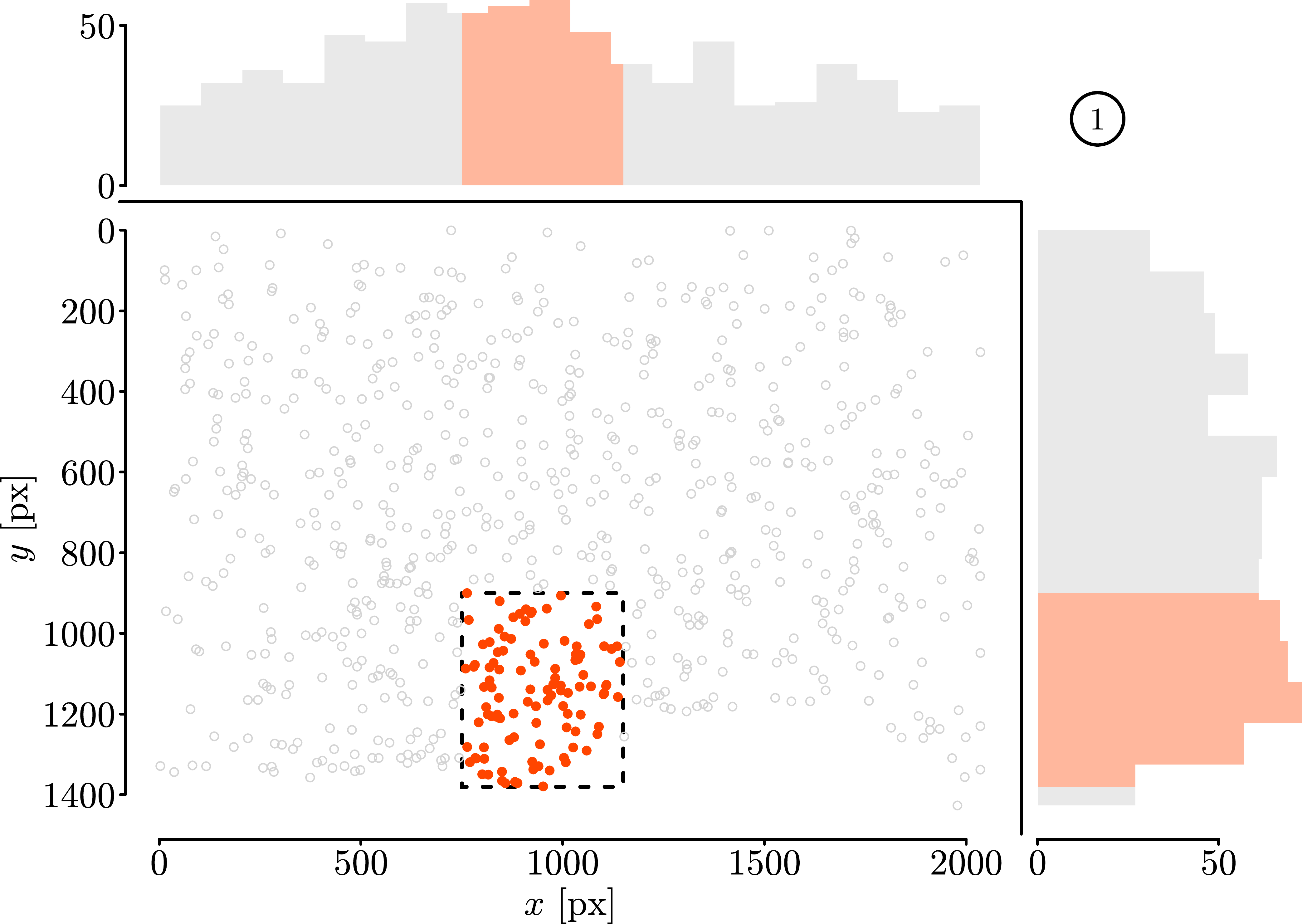}}\vspace{6.5pt}
                \subfloat{\includegraphics[width=\linewidth]{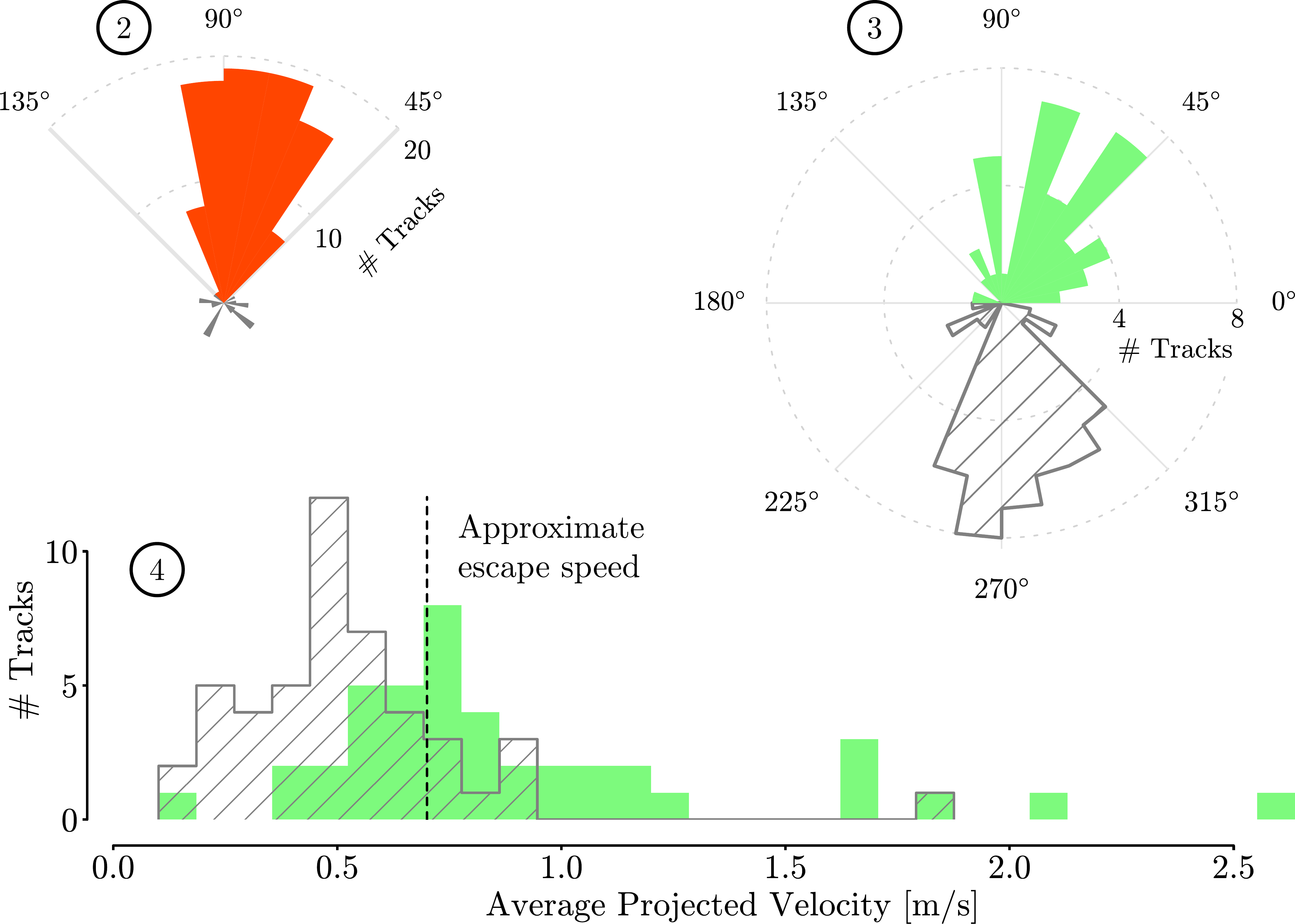}}\vspace{6.5pt}
                \subfloat{\includegraphics[width=\linewidth]{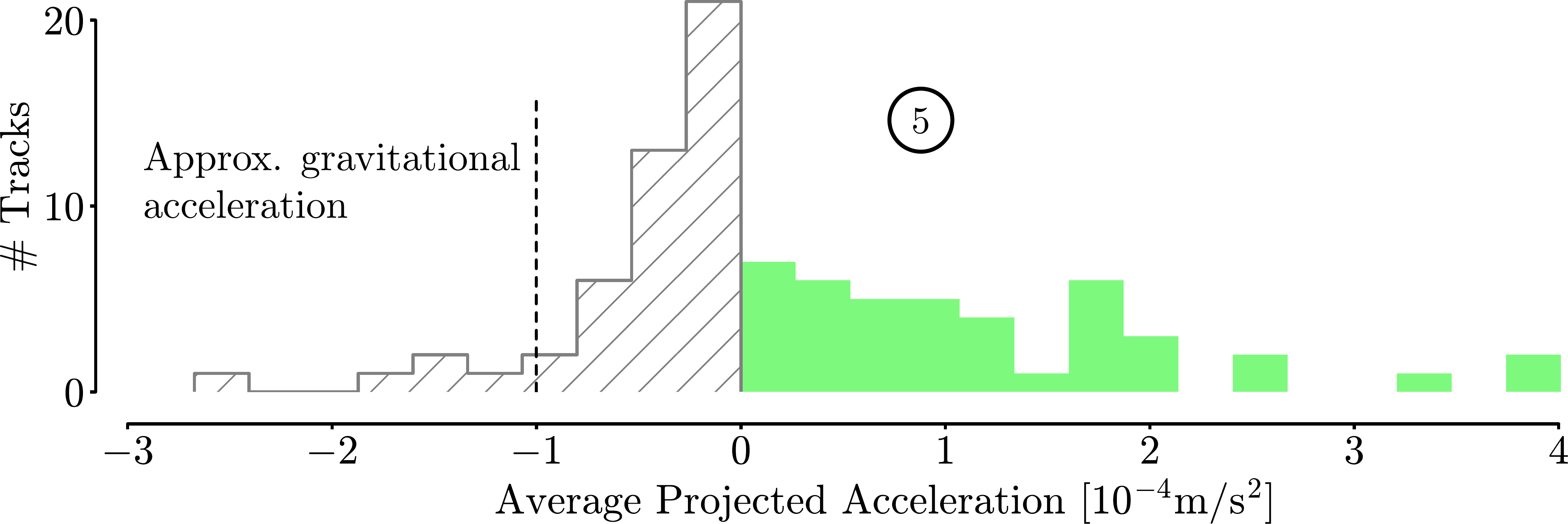}}\vspace{6.5pt}
                \subfloat{\includegraphics[width=\linewidth]{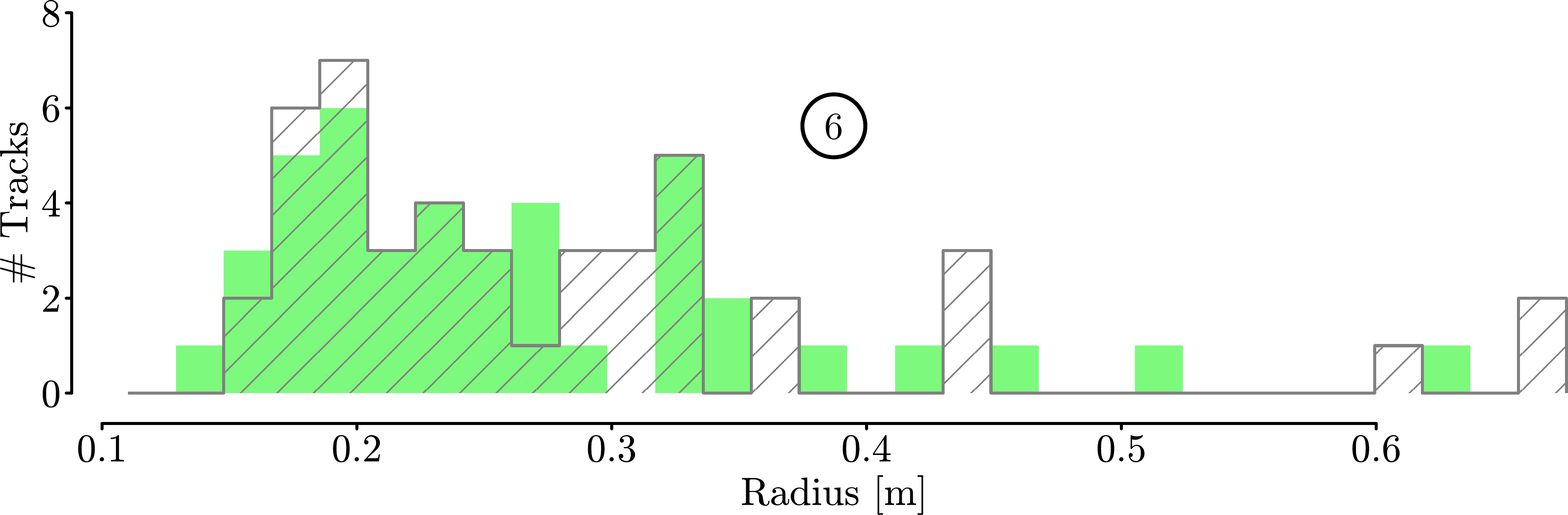}}
                \caption{Selection process and statistics of particles that likely originated from the central active area: (1) the starting points of all 775 tracks (i.e., their earliest confirmed locations) and the tracks we selected (orange circles) that start near the active area. (2) a further reduction of the tracks selected in (1) by choosing only the ones directed upward \mbox{$\pm 45\degree$} (orange). (3) the acceleration angle distribution of the tracks selected in (2), which is further divided into tracks that are accelerated upward (green) and downward (gray hashed). (4, 5, 6) the projected velocity, magnitude of acceleration and radius distributions for the two track populations defined in (3). Escape speed and gravitational acceleration based on \cite{2016Natur.530...63P}.}
                \label{fig:selection_stats}
            \end{figure}
    	    			
			\begin{figure}[!ht]
	  			\centering
				\includegraphics[width=\linewidth]{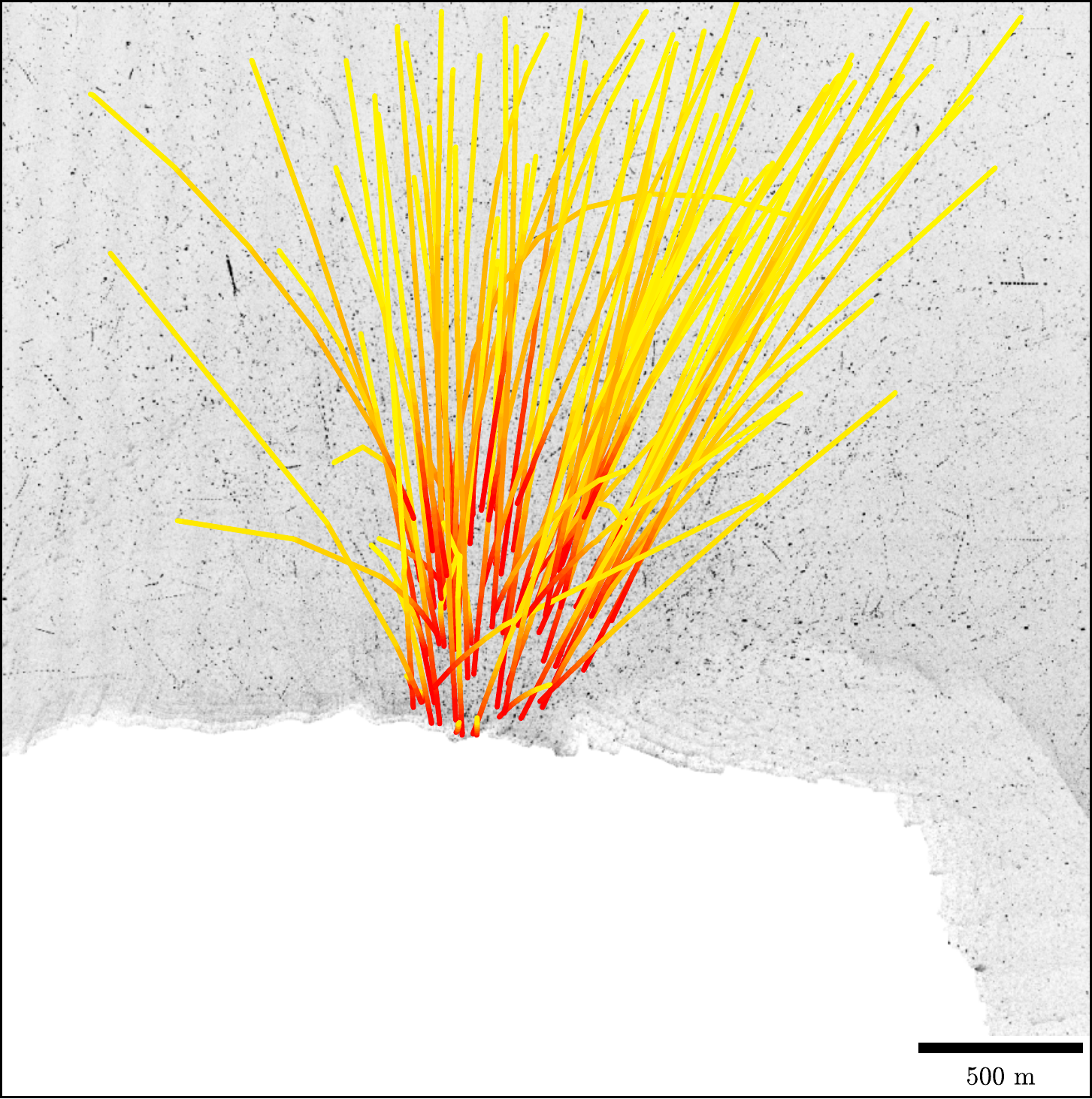}
				\caption{The 89 selected tracks near the central active area.}
				\label{fig:selection_tracks}
			\end{figure}
			
			To derive particle radii and convert particle velocities and accelerations to physical units (e.g., from px/s to m/s), we need to know the particle distances to the spacecraft. Since the only accurate distance measurement we have is of the nucleus ($\sim$86\,km), we focus on particles which were seemingly just ejected from the active area in the center of the images (at that distance \mbox{1\,px $\equiv$ $\sim$\,1.6\,m}). In the following example, we isolated this group in two steps. First, we chose a region around the active area and selected the tracks that originate within it (106 tracks, Fig.~\hyperref[fig:selection_stats]{\ref*{fig:selection_stats}.1}); then, we further reduced the group by selecting only tracks whose projected velocities are pointing upward within \mbox{$\pm 45\degree$} (89 tracks, Fig.~\hyperref[fig:selection_stats]{\ref*{fig:selection_stats}.2}). Figure~\ref{fig:selection_tracks} shows the selected tracks as they appear in front of the stacked image of sequence STP090.
			
			Figure~\hyperref[fig:selection_stats]{\ref*{fig:selection_stats}.3} shows that roughly 47\,\% of the selected tracks have a projected acceleration that points away from the nucleus. The velocity distribution (Fig.~\hyperref[fig:selection_stats]{\ref*{fig:selection_stats}.4}) shows that they are on average faster than the particles that are accelerated downward. Most particles show a net acceleration less negative than gravity (Fig.~\hyperref[fig:selection_stats]{\ref*{fig:selection_stats}.5}). Assuming that on a first order the gravitational acceleration is comparable for all particles they must be experiencing an upward directed acceleration of variable strength that partially compensates or even exceeds gravity. A likely candidate for this upward force is gas drag.
			
			Figure~\hyperref[fig:selection_stats]{\ref*{fig:selection_stats}.6} shows the distribution of the particle radii, which were calculated as:
			\begin{align}
			    \label{eq:radius} r = \sqrt{J \frac{r_\text{h}^2 \Delta^2}{R I_\odot}},
			\end{align}
			where $r$ is the radius in m, $J$ the average particle flux in W/m$^2$/nm, $r_\text{h}$ the dimensionless heliocentric distance measured in units of AU, $\Delta$ the observer-particle distance in m, \mbox{$R=0.0021$} the particle reflectance (computed for decimeter-sized particles using the model in \citealp{8663e156a2a14932bb66452963913cec}), and \mbox{$I_\odot=1.565$\,W/m$^2$/nm} the solar flux in the NAC F22 filter at 1\,AU. The distribution agrees with \cite{10.1093/mnras/stw2179}'s findings when considering that their calculation is affected by a numerical error that leads them to systematically underestimate the radii by a factor of 4.4 (Agarwal et al., in preparation). It furthermore shows no clear trend between the upward- and downward-accelerated particles, which is remarkable because the gas drag we deem responsible for the upward-acceleration should be stronger for smaller particles. 
			
			If we assume that the particles have the same bulk density as the nucleus (533\,kg/m$^3$, \citealp{2016Natur.530...63P}), then the upward-accelerated particles contain about 1300\,kg, while the downward-accelerated ones contain roughly 3000\,kg. The largest boulder alone contains more than half of the mass of the upward-accelerated particles.
			
			A similar extended analysis would be interesting to compare to typical models of cometary dust size distributions \citep[e.g.,][and references therein]{10.1093/mnras/stx2741}, as they predict that the majority of mass lost due to refractory material is likely contained in the largest specimen. Hence knowing the size limit and emission rate of the largest chunks is crucial to estimate a comet's contribution to the interplanetary dust environment and the zodiacal cloud \citep{nesvorny-janches2011}. 
	
	        Lastly, we can also extrapolate our tracks back in time to find out when and where the particles were likely ejected. As the process behind lifting decimeter-sized debris from the surface is not entirely understood (although it appears now to be more straightforward to explain than the lifting of smaller, micron-sized dust, \citealp{gundlach-blum2015}), this can provide us with possible clues about the lifting mechanism or its conditions.

	\section{Summary and outlook}
	\label{sec:conclusions}

	    In this paper we present our algorithm for tracking the motion of debris near the nucleus of comet \object{67P}. The algorithm operates on image sequences recorded by Rosetta's camera system OSIRIS. The sequences typically show part of \object{67P}'s surface that ideally has at least one clearly discernible active area which is ejecting particles that appear as point sources against the dark backdrop of interplanetary space. 
	    
	    As an example and to assess the algorithm's reliability as well as presenting tentative first results, we applied our algorithm to image sequence STP090. The evaluation not only showed that our algorithm can find a large number of tracks, but also revealed a robust criterion--having a miss rate \mbox{$\Gamma<30\,\%$}--to separate genuine from ambiguous tracks. Our first results from a group of particles that satisfied the criterion and likely originated from the central area in sequence STP090 demonstrate one way of how our tracking results can be used. And finally, knowing the projected particle velocities and accelerations can help us estimate the fall-back fraction and the refractory-to-ice ratio--which are key to understanding more about cometary interiors and the role comets play in planetesimal formation.

	\begin{acknowledgements}
	    We thank Eberhard Bodenschatz, Ulrich Christensen, and Pablo Lemos for our fruitful discussions; Carsten Güttler, Michael Mommert, and Jakob Deller for their early support; Kyle Barbary, Benne Holwerda, Peter Stetson, Gábor Kovács, Cecilia Tubiana, Guus Bertens, Jan Molacek, and Maurizio Berti for their technical support; Asmus Freytag for proofreading; and Steve Chesley for reviewing our paper and providing constructive comments. 
	    \newline
	    We acknowledge the operation and calibration team at MPS and the Principal Investigator Holger Sierks on behalf of the OSIRIS Team for providing the OSIRIS images and related datasets. OSIRIS was built by a consortium of the Max-Planck-Institut für Sonnensystemforschung, Göttingen, Germany; the CISAS University of Padova, Italy; the Laboratoire d'Astrophysique de Marseille, France; the Instituto de Astrofísica de Andalucia, CSIC, Granada, Spain; the Research and Scientific Support Department of the European Space Agency, Noordwijk, The Netherlands; the Instituto Nacional de Técnica Aeroespacial, Madrid, Spain; the Universidad Politéchnica de Madrid, Spain; the Department of Physics and Astronomy of Uppsala University, Sweden; and the Institut für Datentechnik und Kommunikationsnetze der Technischen Universität Braunschweig, Germany. The support of the national funding agencies of Germany (DLR), France (CNES), Italy (ASI), Spain (MEC), Sweden (SNSB), and the ESA Technical Directorate is gratefully acknowledged. We thank the Rosetta Science Ground Segment at ESAC, the Rosetta Missions Operations Centre at ESOC and the Rosetta Project at ESTEC for their outstanding work enabling the science return of the Rosetta Mission.
		\newline
		MP and JA acknowledge funding by the ERC Starting Grant No. 757390 Comet and Asteroid Re-Shaping through Activity (CAstRA). JA acknowledges funding by the Volkswagen Foundation.
	\end{acknowledgements}
	
	\bibliographystyle{aa}
	\bibliography{bibl}
	
	\begin{appendix}

		\section{Tracking parameter values}
			
    	    \ctable[
                caption = Tracking parameter values used for sequence STP090.,
                label   = tab:parameters,
                pos     = ht,
                width   = \linewidth,
                notespar
            ]{lcc}{
                Dashes indicate when the parameter values are the same for both principal and extended tracking, $\emptyset$ when the parameter is not used in the given mode. 
            }{                                                                      \FL
                Parameter                           & Principal tracking       & Extended tracking  \ML
                \multicolumn{3}{c}{Educated Guesses}\vspace{3pt}                               \NN
            	$R_\text{init}$ [px]	            & 12                    & --    \NN
            	$R_\text{off}$ [px]	                & 2                     & --    \NN
            	$N_\text{det}$	                    & 8                     & $\emptyset$     \NN
            	$N_\text{pair}$	                    & 3                     & $\emptyset$     \NN
            	Lives	                            & $\emptyset$           & 2\vspace{8pt}     \NN
                \multicolumn{3}{c}{Systematic Optimization}\vspace{3pt}                         \NN
            	$R_\text{max}$ [px]	                & 6.5                   & 9.5   \NN 
            	$R_\text{min}$ [px]	                & 1.7                   & --    \NN 
            	$\hat{d}$ [px]	                    & 1.5                   & --    \NN 
            	$\bar{d}$       	                & 7.8                   & --\vspace{8pt}    \NN
            	$\Omega_\text{max}$ [\textdegree]	& 290                   & 210   \NN 
            	$\Omega_\text{min}$ [\textdegree]	& 40                    & --    \NN 
            	$\hat{d}_\Omega$ [\textdegree]      & 4.4                   & --    \NN 
            	$\bar{d}_\Omega$       	            & 2.0                   & --\vspace{8pt}    \NN
            	$\Delta V_\text{max}$ [\%]  	    & 700                   & 500   \NN 
            	$\Delta V_\text{min}$ [\%]	        & 100                   & --    \NN 
            	$\hat{v}$ [px/s]	                & 0.8                   & --    \NN 
            	$\bar{v}$       	                & 0.8                   & --    \ML
            }

	\end{appendix}

\end{document}